\documentclass[seceq]{ptptex}
\usepackage{wrapft}
\usepackage{graphicx}

\newcommand{\bra}[1]{\langle {#1} |}     %%
\newcommand{\ket}[1]{| {#1} \rangle}     %%
\newcommand{\rket}[1]{| {#1} )}     %%
\newcommand{\wtilde}[1]{\widetilde{#1}} %%

\def\beq{\begin{eqnarray}}
\def\eeq{\end{eqnarray}}
\def\bsub{\begin{subequations}}
\def\esub{\end{subequations}}
\def\b{\begin{equation}}
\def\bs{\begin{split}}
\def\es{\end{split}}
\def\e{\end{equation}}
\begin{document}

\title{Beyond the Schwinger boson representation of the $su(2)$-algebra. I
}
\subtitle{New boson representation based on the $su(1,1)$-algebra and its related problems with application}

\author{%       %Use \sc for the family name
Yasuhiko {\sc Tsue},$^{1}$, Constan\c{c}a {\sc Provid\^encia},$^{2}$
Jo\~ao da {\sc Provid\^encia},$^{2}$  and Masatoshi {\sc Yamamura}$^{3}$
}

\inst{%         %Affiliation, neglected when [addenda] or [errata]
$^{1}$Physics Division, Faculty of Science, Kochi University, Kochi 780-8520, Japan\\
$^{2}$Departamento de F\'{i}sica, Universidade de Coimbra, 3004-516 Coimbra, 
Portugal\\
$^{3}$Department of Pure and Applied Physics, 
Faculty of Engineering Science, Kansai University, Suita 564-8680, Japan
}

\abst{
With the use of two kinds of boson operators, a new boson representation of 
the $su(2)$-algebra is proposed. 
The basic idea comes from the pseudo $su(1,1)$-algebra recently given by the present authors. 
It forms a striking contrast to the Schwinger boson representation of the $su(2)$-algebra 
which is also based on two kinds of bosons. 
This representation may be suitable for describing time-dependence of the 
system interacting with the external environment in the framework of the 
thermo field dynamics formalism, i.e., the phase space doubling. 
Further, several deformations related to the $su(2)$-algebra in this boson representation 
are discussed. 
On the basis of these deformed algebra, various types of time-evolution of a simple boson 
system are investigated. 
}

%\subjectindex{xxxx, xxx}

\maketitle

\section{Introduction}

It may be hardly necessary to mention, but the $su(2)$-algebra has made a central contribution to the 
development of microscopic study of nuclear structure. 
The BCS-Bogoliubov theory may be its typical example. 
We know that the $su(2)$-algebra may be the simplest and the most popular Lie algebra in nuclear structure theory. 
In many cases, the Lie algebras have been treated under the name of boson realizations of Lie algebras. 
The prototype has been called as boson expansion or boson mapping for many-fermion systems, 
which is related to the $so(2N)$-algebra for a space of $N$ single-particle states \cite{1}. 
With the development of the study of the boson expansion, it became to be called the boson realization of the 
Lie algebra \cite{2}. 
The simplest case is the boson realization of the $su(2)$-algebra, which is called as 
the Holstein-Primakoff representation \cite{3}. 
The three generators are expressed in terms of one kind of boson operator for a given 
value of the magnitude of the $su(2)$-spin, which determines the irreducible representation. 
Of course, it is positive-definite. 
We know also another boson representation of the $su(2)$-algebra, that is, 
the Schwinger boson representation \cite{4}. 
The explicit form will be given in the relation (\ref{3-1}) and (\ref{3-2}). 
It consists of two kinds of boson operators. 
Different from the Holstein-Primakoff representation, the magnitude of the $su(2)$-spin is expressed in terms of 
an operator, i.e., half of the sum of two boson number operators, which can be seen in the relation (\ref{3-2b}). 
Clearly, it is a positive-definite operator. 
On the other hand, we know the boson representation of the $su(1,1)$-algebra initiated also by Schwinger \cite{4}. 
This representation is also constructed in terms of two kinds of boson operators. 
The explicit form will be presented in the relations (\ref{2-1}) and (\ref{2-3}). 
In contrast with the case of the $su(2)$-algebra, in this representation, the quantum numbers 
specifying the irreducible representation are the eigenvalues of an operator which is related to half of the difference 
of the two boson number operators, which can be seen in the relation (\ref{2-3b}). 
Therefore, this operator is not positive-definite. 
On the analogy with the magnitude of the $su(2)$-spin, we will call the positive eigenvalue as the magnitude of the $su(1,1)$-spin.

The Schwinger boson representation of the $su(1,1)$-algebra may be not familiar to the field 
of nuclear structure theory mainly treating the zero temperature. 
A merit of this representation is to be able to describe damping and amplifying of an isolated harmonic oscillator induced classical 
mechanically by the velocity-dependent force quantum mechanically in conservative form \cite{5,6,7}. 
In the background of this description, there exists the idea of the thermo field dynamics formalism based on the phase space doubling \cite{8}. 
We should note the following: 
The above-mentioned isolated oscillator is of one dimension and the Schwinger boson representation is constructed in 
two dimensional space, because of the use of two kinds of bosons. 
Therefore, the idea of the phase space doubling is conjectured to be useful in the present problem. 
The original intrinsic oscillator is isolated and it is expressed in terms of one kind of boson and external environment 
in terms of another kind of boson, in which the frequency 
is the same as that of the original one. 
The interaction between both systems is naturally introduced.

If we follow the idea of the phase space doubling, the total Hamiltonian for the present system, ${\hat H}$, can be expressed in the form 
\bsub\label{1-1}
\beq
& &{\hat H}={\hat H}_0+{\hat V}_i\ , 
\label{1-1a}\\
& &{\hat H}_0={\hat H}_{\rm intr}-{\hat H}_{\rm extr}\ . 
\label{1-1b}
\eeq
\esub
Here, ${\hat H}_{\rm intr}$ and ${\hat H}_{\rm extr}$ denote the Hamiltonians of the intrinsic and the external environment system, respectively. 
Both Hamiltonians are of the harmonic oscillator type with the same frequencies. 
The part ${\hat V}_i$ denotes the interaction between both systems. 
It should be noted that ${\hat H}$ is not the total energy operator but the operator for time-evolution 
of the system. 
As for ${\hat V}_i$, a certain linear combination of the raising and the lowering operator of the $su(1,1)$-algebra in the 
Schwinger boson representation was adopted in Refs.\citen{5,6,7}. 
Further, we notice that ${\hat H}_0$ is a constant of motion, because ${\hat H}_0$ is related to the magnitude of the $su(1,1)$-spin. 
In Refs.\citen{6,7}, the Hamiltonian (\ref{1-1}) was treated in the framework of the time-dependent variational 
method. 
Of course, under a careful consideration on the magnitude of the $su(1,1)$-spin as a constant of motion 
in the $su(1,1)$-algebra, the trial state is constructed. 
Therefore, the expectation value of ${\hat H}_0$ does not depend on time, but the expectation value of ${\hat H}_{\rm intr}$ depends on time. 
Under the use of ${\hat V}_i$ mentioned above, 
the expectation value becomes decreasing or increasing function of time. 
The former and the latter show the damped and the amplified oscillation, respectively. 
Some variations of the above-mentioned idea were discussed in Ref.\citen{7}.

However, we must point out that the $su(1,1)$-algebra in the Schwinger boson representation cannot be applied directly 
to many-fermion system. 
The reason is very simple: 
We cannot find the $su(1,1)$-algebra in many-fermion system. 
In response to this situation, the present authors proposed an idea \cite{9}. 
Hereafter, Ref.\citen{9} will be referred to as (A). 
The orthogonal set constructed under the Schwinger boson representation of the $su(1,1)$-algebra 
forms an infinite dimensional space for a given value of the magnitude of the $su(1,1)$-spin. 
But, the orthogonal set for many-fermion system is of finite dimension for a given value 
of the magnitude of the $su(2)$-spin. 
Then, a possible idea is to define, in the infinite dimensional space, a certain subspace which is in one-to-one correspondence to the fermion space. 
Further, we required that any matrix element of the raising and the lowering operator in this subspace does not change 
the form from that given in this algebra. 
If restricted to this subspace, the algebra becomes deformed from the $su(1,1)$-algebra. 
In Ref.\citen{9}, we call it 
the pseudo $su(1,1)$-algebra. 
Three generators are expressed in terms of the original $su(1,1)$-generators with certain parameter which is closely related with 
the dimension of the subspace. 
Then, we can connect the pseudo $su(1,1)$-algebra with the $su(2)$-algebra, for example, which governs 
the Cooper pair. 
The above idea suggests us that many-fermion system can be treated by the pseudo $su(1,1)$-algebra and in (A), 
we described a simple fermion system based on the thermo field dynamics formalism. 
As a result, the periodical dependence of the energy of the intrinsic system on time was shown. 
This result is in contrast to that in the $su(1,1)$-algebra. 
For the time-dependent variational method adopted in (A), we prepare a trial state which contains one complex parameter 
for the variation and the normalization constant of the trial state must be given. 
But, in the case of the pseudo $su(1,1)$-algebra, the explicit form of the normalization constant is too complicated to treat it practically. 
It was shown in (A).

Main aim of this paper is to present a new boson representation of the $su(2)$-algebra which may be 
suitable for the application of the idea of the phase space doubling. 
As was already mentioned, the Schwinger representation is powerless to the use of the phase space doubling. 
In contrast to the case of the Schwinger boson representation, the operator for the magnitude of the $su(2)$-spin in the present case is 
expressed in terms of the form related to the difference of two boson number operators. 
Under an idea similar to that for constructing the pseudo $su(1,1)$-algebra, we can 
express the $su(2)$-generators as functions of the three $su(1,1)$-generators and a certain parameter. 
If the value of this parameter is appropriately chosen, the present representation is reduced to the boson 
realization of the $su(2)$-algebra governing, for example, the Cooper pair. 
Of course, these generators are available in the subspace leading the pseudo $su(1,1)$-algebra. 
In Part II, we will give the proof. 
We intend to apply the present representation to the Hamiltonian (\ref{1-1}) under the same manner as that in (A). 
Interaction term ${\hat V}_i$ must be appropriately chosen. 
Therefore, we prepare a certain trial state for the time-dependent variation so as to calculate the normalization constant easily. 
Expecting various results, we give several types of the deformations from the new boson representation of the $su(2)$-algebra. 
As was already mentioned, the algebra discussed in this paper is applied to the Hamiltonian (\ref{1-1}). 
As for ${\hat V}_i$, we adopt a certain linear combination of the raising and the lowering operator of the algebra obtained by each 
deformation and we can show that the expectation value of the intrinsic Hamiltonian, ${\hat H}_{\rm intr}$, changes periodically for time. 
Depending on the choice of the deformation, the features of the change show various shapes. 
In classical mechanics, we know one problem: 
Elastic collision of simply oscillating light particle with the sufficient heavy one on the 
horizontal straight line. 
We can show that the problem discussed in this paper reduces to the same as the above.

After, in \S 2, recapitulating the pseudo $su(1,1)$-algebra presented in (A) with some new aspects, 
a new boson representation of the $su(2)$-algebra is proposed in \S 3. 
It is based on the Schwinger boson representation of the $su(1,1)$-algebra. 
The proof is given in Part II. 
Section 4 is devoted to giving various deformation from the new boson representation. 
Normalization constant in the orthogonal set obtained in each deformation is easily calculated with rather simple 
approximation. 
In \S\S 5 and 6, the time-evolution is investigated for the boson system 
characterized by the Hamiltonian (\ref{1-1}). 
Depending chosen deformation, the change of the energy of the intrinsic system for time is periodic, but, 
the behaviors are different from one another.

\setcounter{equation}{0}

\section{Pseudo $su(1,1)$-algebra in the Schwinger boson representation}

Our preliminary task is to recapitulate the Schwinger boson representation of the 
$su(1,1)$-algebra in a form suitable for later discussion. 
The details can be seen in Refs.\citen{4,5,6,7}. 
It starts in the following three operators:
\beq\label{2-1}
{\hat T}_+={\hat a}^*{\hat b}^*\ , \qquad
{\hat T}_-={\hat b}{\hat a}\ , \qquad
{\hat T}_0=\frac{1}{2}({\hat a}^*{\hat a}+{\hat b}^*{\hat b}+1)\ . 
\eeq
Here, $({\hat a},{\hat a}^*)$ and $({\hat b},{\hat b}^*)$ denote two kinds of the boson operators. 
The operators ${\hat T}_{\pm,0}$ form the $su(1,1)$-algebra: 
\bsub\label{2-2}
\beq
& &{\hat T}_-^*={\hat T}_+\ , \qquad {\hat T}_0^*={\hat T}_0\ , 
\label{2-2a}\\
& &[\ {\hat T}_+\ , \ {\hat T}_-\ ]=-2{\hat T}_0\ , \qquad
[\ {\hat T}_0\ , \ {\hat T}_{\pm}\ ]=\pm{\hat T}_{\pm}\ . 
\label{2-2b}
\eeq
\esub
The Casimir operator ${\hat {\mib T}}^2$, together with its properties, is given by 
\bsub\label{2-3}
\beq
& &{\hat {\mib T}}^2={\hat T}_0^2-\frac{1}{2}\left({\hat T}_-{\hat T}_+ 
+{\hat T}_+{\hat T}_-\right)={\hat T}\left({\hat T}-1\right)\ , 
\label{2-3a}\\
& &{\hat T}=\frac{1}{2}(-{\hat a}^*{\hat a}+{\hat b}^*{\hat b}+1)\ , \qquad
[\ {\hat T}\ , \ {\hat T}_{\pm,0}\ ]=0\ . 
\label{2-3b}
\eeq
\esub
The eigenstate of ${\hat T}$ and ${\hat T}_0$ with its eigenvalue $t$ and $t_0$, respectively, 
is obtained in the form
\beq\label{2-4}
& &\ket{t,t_0}=\frac{1}{\sqrt{(t_0-t)!(t_0+t-1)!}}
({\hat a}^*)^{t_0-t}({\hat b}^*)^{t_0+t-1}\ket{0}\ , \quad ({\rm normalized})
\nonumber\\
& &{\hat a}\ket{0}={\hat b}\ket{0}=0\ . 
\eeq

Since $t_0-t=0,\ 1,\ 2,\ 3,\cdots$ and $t_0+t-1=0,\ 1,\ 2,\ 3,\cdots$, 
$\ket{t,t_0}$ is treated separately in the following two cases: 
\bsub\label{2-5}
\beq
& &{\rm (i)}\ \ t=1/2,\ 1,\ 3/2, \cdots ,\ \infty\ , \qquad
t_0=t,\ t+1,\ t+2,\cdots ,\ \infty \ \ {\rm for\ a\ given}\ t\ , 
\label{2-5a}\\
& &{\rm (ii)}\ t=0,\ -1/2,\ -1,\cdots ,\ -\infty\ , \quad
t_0=-t+1,\ -t+2,\ -t+3,\cdots ,\ \infty \ \ {\rm for\ a\ given}\ t\ . \nonumber\\
& &\label{2-5b}
\eeq
\esub
It is noted that, in this paper, only the case (i) will be treated. 
In Part II, we will contact with the case (ii) briefly in connection with the case (i). 
The state $\ket{t,t_0}$ can be rewritten in the form
\bsub\label{2-6}
\beq
& &\ket{t,t_0}=\frac{1}{\sqrt{(t_0-t)!(t_0+t-1)!}}\left({\hat T}_+\right)^{t_0-t}\ket{t}\ . 
\quad
(\ket{t,t_0=t}=\ket{t})
\label{2-6a}\\
{\rm i.e.}
& &\ket{n;t}=\frac{1}{\sqrt{n!(2t-1+n)!}}\left({\hat T}_+\right)^{n}\ket{t}\ .
\label{2-6b} 
\eeq
\esub
Here, $\ket{t}$ denotes the minimum weight state:
\beq\label{2-7}
\ket{t}=({\hat b}^*)^{2t-1}\ket{0}\ , \qquad
{\hat T}_-\ket{t}=0\ , \qquad
{\hat T}_0\ket{t}={\hat T}\ket{t}=t\ket{t}\ . 
\eeq
By $n$-time operation of the raising operator ${\hat T}_+$ on $\ket{t}$, $\ket{n;t}$ 
is obtained. 
It should be noted that this operation is permitted until the infinite 
time, in other word, there does not exist the maximum weight state. 
Clearly, $\{$the states (\ref{2-6})$\}$ forms an orthogonal set with the infinite dimension.

Recently, a possible form of the pseudo $su(1,1)$-algebra was presented 
by the present authors \cite{9}. 
It aims at demonstrating a deformation of the Cooper pair obeying the $su(2)$-algebra 
in many-fermion system. 
Hereafter, it will be referred to as (A). 
In (A), one type of possible deformations of the $su(1,1)$-algebra in the 
Schwinger boson representation was treated and we called it the pseudo $su(1,1)$-algebra. 
In this paper, this algebra is formulated in a manner slightly modified form 
that given in (A). 
Basic scheme of the pseudo $su(1,1)$-algebra is to construct it in the 
subspace of the space (\ref{2-5a}):
\beq\label{2-8}
& &t=1/2,\ 1,\ 3/2,\cdots ,\ \mu-1/2,\ \mu\ , \quad
t_0=t,\ t+1,\ t+2,\cdots ,\ t_m-1, \ t_m\nonumber\\
& & \qquad\qquad\qquad\qquad\qquad\qquad\qquad\qquad\qquad\qquad\qquad\qquad  
{\rm for\ a\ given}\ t\ .
\eeq
Here, $\mu$ and $t_m$ denote integer or half-integer, where $t_m$ is a function of $t$. 
Depending on the model under investigation, the value of $\mu$ and the 
functional form of $t_m$ are chosen appropriately. 
Later, we will show a possible idea for the determination of $\mu$ and $t_m$.

Let ${\hat {\cal T}}_{\pm,0}$ denote three generators of the pseudo $su(1,1)$-algebra. 
Role of ${\hat {\cal T}}_{\pm,0}$ is the same as that of ${\hat T}_{\pm,0}$. 
One time operation of ${\hat {\cal T}}_+$ makes the eigenvalue of ${\hat {\cal T}}_0$ 
increase by one in the state specified by the quantum number $(t,t_0)$. 
In this algebra, there exist not only the minimum but also the maximum weight state. 
Further, the following is required for a given $t$: 
the minimum and the maximum weight state are identical to $\ket{t,t}(=\ket{t})$ and 
$\ket{t,t_m}$, respectively, and successive operation of ${\hat {\cal T}}_+$ on $\ket{t}$ 
reduces to the state (\ref{2-6}). 
The above requirement is formulated as follows:
\bsub\label{2-9}
\beq
& &{\hat {\cal T}}_-\ket{t,t}=0\ , \qquad 
{\hat {\cal T}}_0\ket{t,t}=t\ket{t,t}\ , 
\label{2-9a}\\
& &{\hat {\cal T}}_+\ket{t,t_m}=0\ , \qquad
{\hat {\cal T}}_0\ket{t,t_m}=t_m\ket{t,t_m}\ , 
\label{2-9b}\\
& &\left({\hat {\cal T}}_+\right)^{t_0-t}\ket{t}=\left({\hat T}_+\right)^{t_0-t}\ket{t}\ . 
\label{2-9c}
\eeq
\esub
As was already mentioned, the eigenvalue of ${\hat {\cal T}}_0$ increases one by one from 
$t$ to $t_m$ and, then, ${\hat {\cal T}}_0$ may be permitted to be of the form
\beq\label{2-10}
{\hat {\cal T}}_0={\hat T}_0+f({\hat T})\ . 
\eeq
The reason is very simple. 
The eigenvalue of ${\hat T}_0$ increases one by one and $f({\hat T})$ does not make 
the eigenvalue of ${\hat T}_0$ change. 
Then, by operating ${\hat {\cal T}}_0$ on the states $\ket{t,t}$ and 
$\ket{t,t_m}$, the following relation is obtained: 
\beq\label{2-11}
t+f(t)=t\ , \qquad t_m+f(t)=t_m\ , \quad {\rm i.e.,}\quad
f(t)=0\ . 
\eeq
Therefore, ${\hat {\cal T}}_0$ is equal to ${\hat T}_0$:
\bsub\label{2-12}
\beq\label{2-12a}
{\hat {\cal T}}_0={\hat T}_0\ . 
\eeq
If noticing the relation ${\hat T}_-\ket{t}=0$ and 
$\sqrt{t_m-{\hat T}_0}\cdot(\sqrt{t_m-{\hat T}_0+\epsilon})^{-1}\ket{t,t_m}
=$\break
$\sqrt{0}/\sqrt{\epsilon}\ket{t,t_m}\rightarrow 0$ $(\epsilon \rightarrow 0)$, 
the requirement (\ref{2-9}) suggests us the following form for ${\hat {\cal T}}_{\pm}$:
\beq
{\hat {\cal T}}_+={\hat T}_+\cdot\sqrt{t_m-{\hat T}_0}
\cdot \left(\sqrt{t_m-{\hat T}_0+\epsilon}\right)^{-1}\!\! , \quad
{\hat {\cal T}}_-=\left(\sqrt{t_m-{\hat T}_0+\epsilon}\right)^{-1}\!\!\!\cdot\!
\sqrt{t_m-{\hat T}_0}\cdot {\hat T}_- . \nonumber\\
& &
\label{2-12b}
\eeq
\esub
Here, as was already shown, $\epsilon$ denotes positive infinitesimal parameter. 
With the use of the commutation relation (\ref{2-2b}), the arrangement of the 
operators can be changed.

It was already mentioned that $t_m$ is a function of $t$, i.e., $t_m=F_m(t)$. 
Depending on the problem under investigation, the form of $F_m(t)$ is fixed. 
If intending to apply the present algebra to the investigation of such 
state as boson coherent state, 
the present one must be formulated in the operator form, that is, 
there do not appear such quanta as $t_0$, $t$ and $t_m$. 
Then, it may be permissible to define an operator ${\hat T}_m=F_m({\hat T})$ and 
if ${\hat T}$ is replaced with $t$ in $F_m({\hat T})$, ${\hat T}_m$ becomes $F_m(t)=t_m$. 
Therefore, ${\hat {\cal T}}_{\pm,0}$ can be expressed as 
\beq\label{2-13}
& &{\hat {\cal T}}_+={\hat T}_+\cdot\sqrt{{\hat T}_m-{\hat T}_0}\cdot
\left(\sqrt{{\hat T}_m-{\hat T}_0+\epsilon}\right)^{-1}\ , \nonumber\\
& &
{\hat {\cal T}}_-=\left(\sqrt{{\hat T}_m-{\hat T}_0+\epsilon}\right)^{-1}\!\!\!\cdot\!
\sqrt{{\hat T}_m-{\hat T}_0}\cdot{\hat T}_-\ , 
\nonumber\\
& &{\hat {\cal T}}_0={\hat T}_0\ . 
\eeq
The operator ${\hat {\cal T}}_{\pm,0}$ satisfy 
\bsub\label{2-14}
\beq
& &{\hat {\cal T}}_-^*={\hat {\cal T}}_+\ , \qquad {\hat {\cal T}}_0^*={\hat {\cal T}}_0\ , \nonumber\\
& &[\ {\hat {\cal T}}_+\ , \ {\hat {\cal T}}_-\ ]=-2{\hat {\cal T}}_0+\epsilon
\frac{({\hat T}_0+{\hat T})({\hat T}_0-{\hat T}+1)}{{\hat T}_m-{\hat T}_0+\epsilon}\ , 
\label{2-14a}\\
& &[\ {\hat {\cal T}}_0\ , \ {\hat {\cal T}}_{\pm}\ ]=\pm{\hat {\cal T}}_{\pm}\ . 
\label{2-14b}
\eeq
\esub
The operator ${\hat {\mib {\cal T}}}^2$ corresponding to the Casimir operator ${\hat {\mib T}}^2$ is obtained in the form
\beq\label{2-15}
{\hat {\mib {\cal T}}}^2&=&
{\hat {\cal T}}_0^2-\frac{1}{2}\left({\hat {\cal T}}_-{\hat {\cal T}}_+
+{\hat {\cal T}}_+{\hat {\cal T}}_-\right)\nonumber\\
&=&{\hat {\cal T}}\left({\hat {\cal T}}-1\right)+\frac{1}{2}\epsilon
\frac{({\hat T}_0+{\hat T})({\hat T}_0-{\hat T}+1)}{{\hat T}_m-{\hat T}_0+\epsilon}\ , \qquad
{\hat {\cal T}}={\hat T}\ .
\eeq
If, in second terms on the right-hand sides of the relations (\ref{2-14a}) and (\ref{2-15}) the limiting process 
${\hat T}_m\rightarrow \infty$ proceeds to replacing ${\hat T}_0$ and ${\hat T}$ with the eigenvalues $t_0$ 
and $t$, the above algebra reduces to the $su(1,1)$-algebra. 
This can be seen in the form of the generators (\ref{2-12}) directly. 
For example, in the case ${\hat T}_m=C_m+1-{\hat T}$, ${\hat T}_m\rightarrow \infty$ 
if $C_m\rightarrow \infty$. 
Of course, $C_m$ is a parameter. 
This formula may be understood in the following manner: 
In the state (\ref{2-9c}) for a given $t$, the created bosons consist of $2(t_m-t)\ 
(=2(F_m(t)-t))$ bosons in the $a$-$b$ pair type and $(2t-1)$ bosons 
in the single $b$-boson type. 
In the present scheme, the boson in the single $a$-boson type is forbidden to create. 
In the case $t=1/2$, the created bosons are all in the $a$-$b$ pair type, i.e., the 
number of the created bosons is 2 $(F_m(1/2)-1/2)$. 
The above consideration suggests us the following aspect: 
In the state (\ref{2-9c}), there exist $(2t-1)$ vacancies 
and if $(2t-1)$ $a$-bosons occupy these vacancies, all the created bosons 
come to be in the $a$-$b$ pair type. 
On the basis of the above argument, we require that independent of the value of $t$, 
the number of the created bosons is limited in the present system. 
Then, the following relation may be acceptable:
\bsub\label{2-16add}
\beq
& &2(F_m(t)-t)+2(2t-1)=2\left(
F_m\left(\frac{1}{2}\right)-\frac{1}{2}\right)\ (=2C_m)\ , \nonumber\\
{\rm i.e.,}\quad & &
t_m=C_m+1-t\ . 
\label{2-16a}
\eeq
Since $t_m=F_m(t)$ is decreasing for $t$, there exists a certain point $t=t_c$ 
satisfying $F_m(t_c)=t_c$ and, then, $t_c$ is nothing but $\mu$ ($t_c=\mu$). 
This argument gives us 
\beq\label{2-16b}
\mu=\frac{1}{2}(C_m+1)\ .
\eeq
\esub

%If the relation (\ref{2-16a}) is accepted, further, we have the relation 
%\beq
%\mu \leq \frac{1}{2}(C_m+1)\ . 
%\label{2-16b}
%\eeq
%\esub
%Here, $\mu$ denoted the maximum value of $t$ introduced in the relation (\ref{2-8}). 
%The reason is as follows: 
%Since $t\leq t_m$, we have $t\leq C_m+1-t$, 
%i.e., $t\leq (C_m+1)/2$. 
%The maximum value of $t$, $\mu$, should obey also this inequality 
%and, then, we have the relation (\ref{2-16b}). 
%In next section, we will show the case $\mu=(C_m+1)/2$. 

%The state (\ref{2-9c}) suggests that the maximum number of the created bosons consists of 
%$2(t_m-t)$ bosons in the $a$-$b$ pair type, $(2t-1)$ bosons in the single $b$-boson type and $(2t-1)$ vacancies. 
%The addition of these three must be equal to the total number of the created bosons, $2C_m$, which 
%is permitted in the model under investigation. 
%Then, we have 
%\beq\label{2-16add}
%2(t_m-t)+(2t-1)+(2t-1)=2C_m\ , \qquad {\rm i.e.,}\qquad 
%t_m=C_m+1-t\ .
%\eeq

With the aid of the $su(1,1)$- and the pseudo $su(1,1)$-algebra in the Schwinger boson representation, the role 
of the phase space doubling mentioned qualitatively in \S 1 can be understood 
rather quantitatively. 
From the relations (\ref{2-1}) and (\ref{2-3b}), the following expression for ${\hat b}^*{\hat b}$ is derived:
\bsub\label{2-17}
\beq\label{2-16}
{\hat b}^*{\hat b}={\hat T}_0+{\hat T}-1\ . 
\eeq
Let a time-dependent state vector, which is the eigenstate of ${\hat T}$ with the eigenvalue $t$, exist. 
This is of the superposition of the orthogonal set (\ref{2-4}) or (\ref{2-6}) from $t_0=t$ to $t\rightarrow \infty$. 
Then, the expectation value of ${\hat b}^*{\hat b}$ at the time $\tau$ is expressed as 
\beq\label{2-17b}
\langle {\hat b}^*{\hat b}\rangle_{\tau}=t_0(\tau)+t-1\ , \qquad
t_0(\tau)=\langle {\hat T}_0\rangle_{\tau}\ , \qquad t_0(0)>t\ . 
\eeq
\esub
The damped and amplified harmonic oscillator can be understood by investigating the behavior 
of $t_0(\tau)$, which is determined by the Hamiltonian adopted in the model under investigation. 
If $t_0(\tau)$ is monotone-decreasing $(t_0(0)>t_0(\tau)>t)$, 
$t_0(\tau)\rightarrow t\ (\tau\rightarrow \infty)$. 
Then, $\langle {\hat H}_{\rm intr}\rangle_{\tau}=\langle \omega{\hat b}^*{\hat b}\rangle_{\tau}$ decreases from the 
value $\omega(t_0(\tau)+t-1)$ to $\omega(2t-1)$ at the limit $\tau\rightarrow \infty$. 
This corresponds to the damped oscillator. 
On the other hand, if $t_0(\tau)$ is monotone-increasing $(t_0(0)<t_0(\tau))$, $t_0(\tau)\rightarrow \infty\ (\tau\rightarrow \infty)$. 
Then, $\langle \omega{\hat b}^*{\hat b}\rangle_{\tau}$ increases from the value $\omega(t_0(0)+t-1)$ to $\infty$ at 
the time $\tau\rightarrow \infty$. 
This case corresponds to the amplified oscillation.

It may be interesting to investigate the pseudo $su(1,1)$-algebra in relation to the 
boson realization of many-fermion system. 
In (A), a possible deformation of the Cooper pair in the frame of this algebra 
was discussed. 
Let a time-dependent state, which is the eigenstate of ${\hat T}$ with the eigenvalue $t$, exist.  
In this case, the state is of a superposition of the eigenstates of ${\hat T}_0$, the eigenvalues of which change from $t_0=t$ to 
$t_0=t_m$. 
Therefore, the expectation value of ${\hat b}^*{\hat b}$ for this state at the time $\tau$ is given in the relation (\ref{2-17}), 
but $t_0(\tau)$ should obey the inequality 
\beq\label{2-18}
t \leq t_0(\tau) \leq t_m\ .
\eeq
Since $\langle {\hat b}^*{\hat b}\rangle_{\tau}$ changes in the range (\ref{2-18}), for example, $\langle {\hat b}^*{\hat b}\rangle_{\tau}$ can change periodically in the range 
\beq\label{2-19}
2t-1 \leq \langle {\hat b}^*{\hat b} \rangle_{\tau} \leq t_m+t-1\ . 
\eeq
Of course, it depends on the Hamiltonian. 
In (A), an example of the periodical change was shown. 
In conclusion, in order to make the idea of the phase space doubling effective in the $su(1,1)$- and the pseudo $su(1,1)$-algebra, 
the operator $({\hat b}^*{\hat b}-{\hat a}^*{\hat a})$, i.e., ${\hat T}$ should be a constant of motion.

\setcounter{equation}{0}

\section{A new boson realization of many-fermion system obeying the $su(2)$-algebra in terms of the $su(1,1)$-algebra}

In (A), the pseudo $su(1,1)$-algebra in the fermion space was discussed in terms of a possible deformation of the 
Cooper pair obeying the $su(2)$-algebra. 
In this paper, Part I, the pseudo $su(1,1)$-algebra is treated from the side of the $su(2)$-algebra in the frame of 
the boson space constructed by two kinds of bosons $({\hat a},{\hat a}^*)$ and $({\hat b},{\hat b}^*)$. 
One of the popular boson representations of the $su(2)$-algebra is presented by Schwinger \cite{4}:
\beq\label{3-1}
{\hat S}_+={\hat a}^*{\hat b}\ , \qquad
{\hat S}_-={\hat b}^*{\hat a}\ , \qquad
{\hat S}_0=\frac{1}{2}({\hat a}^*{\hat a}-{\hat b}^*{\hat b})\ . 
\eeq
Here, ${\hat S}_{\pm,0}$ denote the generators. 
The Casimir operator ${\hat {\mib S}}^2$ is given by 
\bsub\label{3-2}
\beq
& &{\hat {\mib S}}^2={\hat S}_0^2+\frac{1}{2}\left({\hat S}_-{\hat S}_+ +{\hat S}_+{\hat S}_-\right)={\hat S}\left({\hat S}+1\right)\ , 
\label{3-2a}\\
& &{\hat S}=\frac{1}{2}({\hat a}^*{\hat a}+{\hat b}^*{\hat b})\ , \qquad 
[\ {\hat S}\ , \ {\hat S}_{\pm,0}\ ]=0\ . 
\label{3-2b}
\eeq
\esub
We note the relation 
\beq\label{3-3}
{\hat S}={\hat T}_0-\frac{1}{2}\ , \qquad
{\hat S}_0=-{\hat T}+\frac{1}{2}\ . 
\eeq
Since ${\hat S}$ is not related to ${\hat T}$, the form (\ref{3-1}) may be 
not suitable for treating the Hamiltonian (\ref{1-1}) based on the phase space doubling, 
even if it is expressed in terms of the $su(2)$-generators. 
Rather, it may be better to apply this representation to the problem of the energy transfer between the $b$- and the $a$-boson system.

The above mentioning suggests us that, in order to obtain a $su(2)$-algebra in the boson representation for the phase space doubling, 
it may be necessary to connect the operator ${\hat S}$ with ${\hat T}$ directly. 
With this aim, the same idea as that in the pseudo $su(1,1)$-algebra (\ref{2-9}) is adopted: 
\bsub\label{3-4}
\beq
& &{\hat {\cal S}}_-\ket{t,t}=0\ , \qquad
{\hat {\cal S}}_0\ket{t,t}=-s\ket{t,t}\ , 
\label{3-4a}\\
& &{\hat {\cal S}}_+\ket{t,t_m}=0\ , \qquad
{\hat {\cal S}}_0\ket{t,t_m}=s\ket{t,t_m}\ , 
\label{3-4b}\\
& &\left({\hat {\cal S}}_+\right)^{t_0-t}\ket{t}=\sqrt{\frac{(2t-1)!}{(t_0+t-1)!}\cdot\frac{(t_m-t)!}{(t_m-t_0)!}}\left({\hat T}_+\right)^{t_0-t}\ket{t}\ . 
\label{3-4c}
\eeq
\esub
Since the eigenvalue of ${\hat {\cal S}}_0$ increases one by one from $-s$ to $s$ and, then, 
${\hat {\cal S}}_0$ may be permitted to set up the relation 
\beq\label{3-5}
{\hat {\cal S}}_0={\hat T}_0+f({\hat T})\ . 
\eeq
The above is the same as the case of the pseudo $su(1,1)$-algebra. 
Operating ${\hat {\cal S}}_0$ on the states $\ket{t,t}$ and $\ket{t,t_m}$, the following 
relation is obtained: 
\beq\label{3-6}
t+f(t)=-s\ , \qquad 
t_m+f(t)=s\ . 
\eeq
The relation (\ref{3-6}) leads us to $f(t)=-(t_m+t)/2$ and $s=(t_m-t)/2$. 
Therefore, ${\hat {\cal S}}_0$ can be expressed as 
\bsub\label{3-7}
\beq\label{3-7a}
{\hat {\cal S}}_0={\hat T}_0-\frac{1}{2}\left({\hat T}_m+{\hat T}\right)\ .
\eeq
Later, we contact with $s=(t_m-t)/2$. 
If noticing the relations ${\hat T}_-\ket{t}=0$ and $\sqrt{{\hat T}_m-{\hat T}_0}\ket{t,t_m}=0$, the requirement 
(\ref{3-4}) suggests the following form for ${\hat {\cal S}}_{\pm}$:
\beq\label{3-7b}
& &{\hat {\cal S}}_+={\hat T}_+\cdot\sqrt{{\hat T}_m-{\hat T}_0}\cdot\!\!\left(\sqrt{{\hat T}_0+{\hat T}+\epsilon}\right)^{-1}\ , \nonumber\\
& &
{\hat {\cal S}}_-=\left(\sqrt{{\hat T}_0+{\hat T}+\epsilon}\right)^{-1}\!\!\cdot\sqrt{{\hat T}_m-{\hat T}_0}\cdot{\hat T}_- \ .
\eeq
\esub
In the space obeying the relation (\ref{3-4}), 
the operators ${\hat {\cal S}}_{\pm,0}$ satisfy 
\bsub\label{3-8}
\beq
& &{\hat {\cal S}}_-^*={\hat {\cal S}}_+\ , \qquad
{\hat {\cal S}}_0^*={\hat {\cal S}}_0\ , 
\label{3-8a}\\
& &[\ {\hat {\cal S}}_+\ , \ {\hat {\cal S}}_-\ ]=2{\hat {\cal S}}_0\ , \qquad
[\ {\hat {\cal S}}_0\ , \ {\hat {\cal S}}_{\pm}\ ]=\pm{\hat {\cal S}}_{\pm}\ . 
\label{3-8b}
\eeq
%\esub
The proof of the relation (\ref{3-8b}) will be shown in Part II. 
Certainly, ${\hat {\cal S}}_{\pm,0}$ form the $su(2)$-algebra. 
The Casimir operator ${\hat {\mib {\cal S}}}^2$ can be expressed as 
\beq\label{3-9}
{\hat {\mib {\cal S}}}^2={\hat {\cal S}}\left({\hat {\cal S}}+1\right)\ , \qquad
{\hat {\cal S}}=\frac{1}{2}\left({\hat T}_m-{\hat T}\right)\ . 
\eeq
\esub
The operator ${\hat {\cal S}}$ is given from the form $s=(t_m-t)/2$ 
obtained in the relation (\ref{3-6}). 
%The operator ${\hat {\cal S}}$ is given in the relation $s=(t_m-t)/2$ obtained in the relation (\ref{3-6}). 
%The above is a new boson representation. 
%The operator ${\hat {\cal S}}$ can be expressed as 
%\beq\label{3-10}
%{\hat {\cal S}}&=&\frac{1}{2}\left({\hat T}_m-\frac{1}{2}(-{\hat a}^*{\hat a}+{\hat b}^*{\hat b}+1)\right) \nonumber\\
%&=&-\frac{1}{4}\left(({\hat b}^*{\hat b}-{\hat a}^*{\hat a})-\left(2{\hat T}_m-1\right)\right)\ . 
%\eeq
If ${\hat T}_m=C_m+1-{\hat T}$, ${\hat {\cal S}}_{\pm,0}$ and ${\hat {\cal S}}$ can be expressed as follows: 
\bsub\label{3-9add}
\beq
& &{\hat {\cal S}}_+={\hat a}^*{\hat b}^*\cdot
\sqrt{C_m-{\hat b}^*{\hat b}}\cdot\left(\sqrt{{\hat b}^*{\hat b}+1+\epsilon}\right)^{-1}\ , 
\label{3-9a}\\
& &{\hat {\cal S}}_-=\left(\sqrt{{\hat b}^*{\hat b}+1+\epsilon}\right)^{-1}\cdot
\sqrt{C_m-{\hat b}^*{\hat b}}\cdot{\hat b}{\hat a}\ , 
\label{3-9b}\\
& &{\hat {\cal S}}_0=\frac{1}{2}({\hat a}^*{\hat a}+{\hat b}^*{\hat b})-\frac{1}{2}C_m\ , 
\label{3-9c}
\eeq
\esub
\beq\label{3-10}
{\hat {\cal S}}=\frac{1}{2}({\hat a}^*{\hat a}-{\hat b}^*{\hat b})+\frac{1}{2}C_m \ .\qquad\ \ 
\eeq
The above is a new boson representation of the $su(2)$-algebra. 
The expressions (\ref{3-7}) and (\ref{3-9}) hold in the subspace (\ref{2-8}) of the 
space (\ref{2-5a}). 
The detail explanation will be given in Part II. 
The form (\ref{3-10}) suggests that the present representation may be suitable for the 
phase space doubling, because ${\hat b}^*{\hat b}$ and ${\hat a}^*{\hat a}$ appear in the relation of the 
subtraction of the $b$- and the $a$-boson number.

We have developed a new boson representation of the $su(2)$-algebra. 
Such boson representations have played a role of describing various gross properties 
of many-fermion systems. 
In these studies, the investigation on the behaviors of individual fermions is of secondary importance. 
These have been called the boson realization of the Lie algebraic approach to many-fermion problems. 
The $su(2)$-pairing model is a typical example. 
In this model, three quantities occupy central part of the gross properties, that is, 
the total number of the single-particle states $4\Omega_0$ (if follows (A), conventionally $2\Omega$), the 
total fermion number $N$ and the seniority number $\nu$. 
Therefore, in order to complete the present new boson representation in relation to the $su(2)$-pairing model, 
it may be inevitable to connect $t$, $t_0$ and $t_m$ with $\nu$, $N$ and $\Omega_0$.

First, the form ${\wtilde S}_0=(1/2){\wtilde N}-\Omega_0$ in the relation (A.3.1) is taken 
up in terms of the eigenvalue $s_0$ and $N$: 
\bsub\label{3-11}
\beq\label{3-11a}
s_0=\frac{1}{2}N-\Omega_0\ . 
\eeq
In the case $s_0=-s$, $-s=N_{\rm min}/2-\Omega_0$ and $N_{\rm min}=\nu$, and then, 
the following relation is derived:
\beq\label{3-1b}
s=\Omega_0-\frac{1}{2}\nu\ . 
\eeq
\esub
Further, in the case $s_0=s$, $s=N_{\rm max}/2-\Omega_0=\Omega_0-\nu/2$, and then, 
$N_{\rm max}=4\Omega_0-\nu$. 
Noticing the relation $[\ {\wtilde N}\ , \ {\wtilde S}_+\ ]=2{\wtilde S}_+$, $N$ can be given as 
\beq\label{3-12}
N=\nu\ , \ \nu+2\ , \cdots , \ 4\Omega_0-\nu\ (=\nu+2(2\Omega_0-\nu))\ . 
\eeq
Since $s\geq 0$, $\nu$ is given as 
\beq\label{3-13}
\nu=0\ , \ 1\ , \ 2\ , \cdots ,\ 2\Omega_0\ .
\eeq

On the other hand, the relations (\ref{3-7a}) and (\ref{3-9}) lead us to 
\bsub\label{3-14}
\beq
& &s_0=t_0-\frac{1}{2}(t_m+t)\ , 
\label{3-14a}\\
& &s=\frac{1}{2}(t_m-t)\ . 
\label{3-14b}
\eeq
\esub
Equating the relations (\ref{3-11}) and (\ref{3-14}) with each other, the following is obtained: 
\bsub\label{3-15}
\beq
& &2t-1=\nu+\frac{1}{2}\left((2t_0-1)-N\right)\ , 
\label{3-15a}\\
& &2t_m-1=(4\Omega_0-\nu)+\frac{1}{2}\left((2t_0-1)-N\right)\ . 
\label{3-15b}
\eeq
\esub
Discussion on the relations (\ref{3-15a}) and (\ref{3-15b}) starts in a postulate mentioned below. 
The boson vacuum $\ket{0}=\ket{t=1/2, t_0=1/2}$ corresponds 
to the fermion vacuum $\rket{0}=\rket{\nu=0, N=0}$. 
This postulate suggests that the relations (\ref{2-8}) and (\ref{2-13}) give us 
\bsub\label{3-16}
\beq\label{3-16a}
2t-1=\nu\ .
\eeq
The relations (\ref{3-15a}) and (\ref{3-15b}) lead us to 
\beq
& &2t_0-1=N\ , 
\label{3-16b}\\
{\rm i.e.,}& &\nonumber\\
& &2t_m-1=4\Omega_0-\nu\ . 
\label{3-16c}
\eeq
\esub
Since $2t_m-1=4\Omega_0-(2t-1)$, the following relation is obtained: 
\beq\label{3-17}
t_m=(2\Omega_0+1)-t\ . 
\eeq
The relations (\ref{3-14b}) and (\ref{3-17}) give 
\bsub\label{3-18}
\beq\label{3-18a}
s=\left(\Omega_0+\frac{1}{2}\right)-t\ .
\eeq
The maximum values of $t$ and $\nu$ are $\mu$ and $2\Omega_0$, respectively, and the relations (\ref{3-13}) and 
(\ref{3-15a}) give us 
\beq\label{3-18b}
2\mu-1=2\Omega_0\ , \qquad {\rm i.e.,}\qquad \mu=\Omega_0+\frac{1}{2}\ .
\eeq
\esub

The above argument presents us the operator form of the $su(2)$-algebra. 
First, the seniority and the total fermion number operator ${\hat \nu}$ and ${\hat N}$, respectively, are introduced in the form 
\bsub\label{3-19}
\beq
& &{\hat \nu}=2{\hat T}-1=-{\hat a}^*{\hat a}+{\hat b}^*{\hat b}\ , 
\label{3-19a}\\
& &{\hat N}=2{\hat T}_0-1={\hat a}^*{\hat a}+{\hat b}^*{\hat b}\ . 
\label{3-19b}
\eeq
\esub
Further, ${\hat T}_m$ is given as 
\beq\label{3-20}
{\hat T}_m=(2\Omega_0+1)-{\hat T}=2\Omega_0+\frac{1}{2}+\frac{1}{2}({\hat a}^*{\hat a}-{\hat b}^*{\hat b})\ . 
\eeq
The operators ${\hat {\cal S}}_{\pm,0}$ and ${\hat {\cal S}}$ can be expressed as 
\bsub\label{3-21}
\beq
{\hat {\cal S}}_+&=&{\hat T}_+\cdot\sqrt{(2\Omega_0+1)-\left({\hat T}_0+{\hat T}\right)}\cdot\left(
\sqrt{{\hat T}_0+{\hat T}+\epsilon}\right)^{-1} \nonumber\\
&=&{\hat a}^*{\hat b}^*\cdot\sqrt{2\Omega_0-{\hat b}^*{\hat b}}\cdot\left(\sqrt{{\hat b}^*{\hat b}+1+\epsilon}\right)^{-1}\ , 
\label{3-21a}\\
{\hat {\cal S}}_-&=&\left(
\sqrt{{\hat T}_0+{\hat T}+\epsilon}\right)^{-1} \cdot\sqrt{(2\Omega_0+1)-\left({\hat T}_0+{\hat T}\right)}\cdot{\hat T}_-\nonumber\\
&=&\left(\sqrt{{\hat b}^*{\hat b}+1+\epsilon}\right)^{-1}\cdot\sqrt{2\Omega_0-{\hat b}^*{\hat b}}\cdot{\hat b}{\hat a}\ , 
\label{3-21b}\\
{\hat {\cal S}}_0&=&{\hat T}_0-\left(\Omega_0+\frac{1}{2}\right)=\frac{1}{2}({\hat a}^*{\hat a}+{\hat b}^*{\hat b})-\Omega_0\ , 
\label{3-21c}
\eeq
\esub
\beq\label{3-22}
{\hat {\cal S}}\ =\ \left(\Omega_0+\frac{1}{2}\right)-{\hat T}=\frac{1}{2}({\hat a}^*{\hat a}-{\hat b}^*{\hat b})+\Omega_0\ . \qquad\quad
\eeq
The above corresponds to the case $C_m=2\Omega_0$ in the relation (\ref{3-7}) and (\ref{3-9add}). 
Needless to say, the expressions (\ref{3-21}) and (\ref{3-22}) hold in the subspace (\ref{2-8}) of 
the space (\ref{2-5a}). 
The detail will be mentioned in Part II. 
The expression (\ref{3-21}) and (\ref{3-22}) form the third boson representation 
of the $su(2)$-algebra. 
Of course, the first and the second are the Holstein-Primakoff and the Schwinger boson representation, respectively \cite{3,4}.

\setcounter{equation}{0}

\section{Various deformations of the $su(2)$-algebra in the third boson representation}

In this section, we will investigate various deformations of the $su(2)$-algebra developed in \S 3. 
For this aim, let us rewrite the generators ${\hat {\cal S}}_{\pm}$ shown in the relation (\ref{3-7b}) as follows: 
\bsub\label{4-1}
\beq
& &{\hat {\cal S}}_+={\hat T}_+\cdot\sqrt{{\hat T}_m-{\hat T}_0}\cdot\sqrt{1+\left({\hat T}_0-{\hat T}\right)}\cdot\left(
\sqrt{{\hat T}_-{\hat T}_++\epsilon}\right)^{-1}\ , 
\label{4-1a}\\
& &{\hat {\cal S}}_-=\left(\sqrt{{\hat T}_-{\hat T}_++\epsilon}\right)^{-1}\cdot\sqrt{1+\left({\hat T}_0-{\hat T}\right)}\cdot
\sqrt{{\hat T}_m-{\hat T}_0}\cdot{\hat T}_-\ . 
\label{4-1b}
\eeq
\esub
The operator ${\hat {\cal S}}_0$ is unchanged from the form (\ref{3-7a}) and ${\hat T}_-{\hat T}_+$ is given as 
\bsub\label{4-2}
\beq
{\hat T}_-{\hat T}_+=\left({\hat T}_0+{\hat T}\right)\left({\hat T}_0-{\hat T}+1\right)
=\left({\hat T}_0+{\hat T}\right)\left(1+\left({\hat T}_0-{\hat T}\right)\right)\ . 
\label{4-2a}
\eeq
If ${\hat T}_0$ is replaced with $({\hat T}_0-1)$, ${\hat T}_-{\hat T}_+$ becomes to ${\hat T}_+{\hat T}_-$:
\beq
{\hat T}_+{\hat T}_-=\left({\hat T}_0-{\hat T}\right)\left({\hat T}_0+{\hat T}-1\right)\ .
\label{4-2b}
\eeq
\esub
With the use of the relation (\ref{4-2a}), it may be easily verified that the form (\ref{4-1}) is equivalent to the 
relation (\ref{3-7b}).

The discussion starts in the introduction of the operator ${\hat {\cal R}}_{\pm}$ deformed from ${\hat {\cal S}}_{\pm}$:
\bsub\label{4-3}
\beq
& &{\hat {\cal R}}_+={\hat T}_+\cdot\sqrt{{\hat T}_m-{\hat T}_0}\cdot\sqrt{{\hat Q}_p+p\left({\hat T}_0-{\hat T}\right)}\cdot\left(
\sqrt{{\hat T}_-{\hat T}_++\epsilon}\right)^{-1}\ , 
\label{4-3a}\\
& &{\hat {\cal R}}_-=\left(\sqrt{{\hat T}_-{\hat T}_++\epsilon}\right)^{-1}\cdot\sqrt{{\hat Q}_p+p\left({\hat T}_0-{\hat T}\right)}\cdot
\sqrt{{\hat T}_m-{\hat T}_0}\cdot{\hat T}_-\ . 
\label{4-3b}
\eeq
\esub
Here, ${\hat Q}_p$ is a function of ${\hat T}$:
\beq\label{4-4}
{\hat Q}_p=Q_p({\hat T})\ , \qquad
{\hat Q}_p\ket{t,t_0}=Q_p(t)\ket{t,t_0}\ , \qquad
Q_p(t)=q_{p,t}\ . 
\eeq
From the outside, we must fix the concrete form of ${\hat Q}_p$ and 
corresponding to the form of ${\hat Q}_p$, the deformation is determined. 
Without loss of generality, the following conditions are added: 
\beq\label{4-5}
p=\pm 1,\ 0\quad {\rm and\ if}\quad p=0\ , \ {\hat Q}_0=1\ , \quad{\rm i.e.,}\quad q_{0,t}=1\ .  
\eeq
The operator $({\hat Q}_p+p({\hat T}_0-{\hat T}))$ should be positive-definite and it can be shown in the form 
\bsub\label{4-6}
\beq
& &\sqrt{{\hat Q}_p+p\left({\hat T}_0-{\hat T}\right)}\ket{t,t}=\sqrt{q_{p,t}}\ket{t,t}\ , 
\label{4-6a}\\
& &\sqrt{{\hat Q}_p+p\left({\hat T}_0-{\hat T}\right)}\ket{t,t_m}=\sqrt{q_{p,t}+p(t_m-t)}\ket{t,t_m}\ . 
\label{4-6b}
\eeq
\esub
The relation (\ref{4-6}) leads us to 
\bsub\label{4-7}
\beq
& &q_{p,t}\geq 0\ , 
\label{4-7a}\\
& &q_{p,t}+p(t_m-t) \geq 0\ , \qquad {\rm i.e.,}\qquad q_{1,t}\geq -(t_m-t)\ , \quad 
q_{-1,t}\geq t_m-t\ . 
\label{4-7b}
\eeq
\esub
If $q_{p,t}=0$, $\sqrt{{\hat Q}_p+p({\hat T}_0-{\hat T})}\ket{t,t}=0$ and, then, 
${\hat {\cal R}}_+\ket{t,t}=0$, which is not 
interesting, because of ${\hat {\cal R}}_-\ket{t,t}=0$. 
Therefore, in the relation (\ref{4-7a}), $q_{p,t}\geq 0$ should be changed to $q_{p,t} >0$. 
From the above argument, $q_{p,t}$ should obey the condition for the positive-definiteness
\beq\label{4-8}
q_{1,t}>0\ , \qquad q_{-1,t} \geq t_m-t\ . 
\eeq
The operators ${\hat {\cal R}}_{\pm}$ are reduced to ${\hat {\cal S}}_{\pm}$ shown in the relation (\ref{4-1}): 
\beq\label{4-9}
{\rm If}\quad {\hat Q}_1=1\ , \qquad {\hat {\cal R}}_{\pm}={\hat {\cal S}}_{\pm}\ . 
\eeq
The products of ${\hat {\cal R}}_+$ and ${\hat {\cal R}}_-$ are expressed as 
\bsub\label{4-10}
\beq
& &{\hat {\cal R}}_+{\hat {\cal R}}_-=\left({\hat T}_m-{\hat T}_0+1\right)
\left({\hat Q}_p+p\left({\hat T}_0-{\hat T}-1\right)\right)\left(1-\frac{\epsilon}{{\hat T}_+{\hat T}_- +\epsilon}\right)\ , 
\label{4-10a}\\
& &{\hat {\cal R}}_-{\hat {\cal R}}_+=\left({\hat T}_m-{\hat T}_0\right)
\left({\hat Q}_p+p\left({\hat T}_0-{\hat T}\right)\right)\left(1-\frac{\epsilon}{{\hat T}_-{\hat T}_+ +\epsilon}\right)\ . 
\label{4-10b}
\eeq
\esub

In the case $p=\pm 1$, the relation (\ref{4-10}) gives us the commutation relation 
\beq\label{4-11}
[\ {\hat {\cal R}}_+\ , \ {\hat {\cal R}}_-\ ]=p\cdot 2{\hat {\cal R}}_0-\left({\hat T}_m-{\hat T}+1\right)\sum_t \ket{t}\bra{t}\ . 
\eeq
Here, ${\hat {\cal R}}_0$ is defined as 
\beq\label{4-12}
{\hat {\cal R}}_0={\hat T}_0-\frac{1}{2}\left({\hat T}_m+{\hat T}-p{\hat Q}_p+1\right)\ , \qquad
[\ {\hat {\cal R}}_0\ , \ {\hat {\cal R}}_{\pm}\ ]=\pm {\hat {\cal R}}_{\pm}\ .
\eeq
Under the condition (\ref{4-9}), ${\hat {\cal R}}_0$ is reduced to ${\hat {\cal S}}_0$:
\beq\label{4-13}
{\rm If}\quad {\hat Q}_1=1\ , \qquad {\hat {\cal R}}_0={\hat {\cal S}}_0\ .
\eeq
The operator ${\hat {\mib {\cal R}}}^2$ corresponding to the Casimir operator ${\hat {\mib {\cal S}}}^2$ is expressed in the form 
\beq\label{4-14}
{\hat {\mib {\cal R}}}^2&=&
{\hat {\cal R}}_0^2+p\cdot \frac{1}{2}\left({\hat {\cal R}}_-{\hat {\cal R}}_+ +{\hat {\cal R}}_+{\hat {\cal R}}_-\right) \nonumber\\
&=&
F_p({\hat {\cal R}})+\frac{1}{2}\left({\hat T}_m-{\hat T}+1\right)\left({\hat Q}_p-p\right)
\sum_t \ket{t}\bra{t}\ .
\eeq
Here, ${\hat {\cal R}}$ and $F_p({\hat {\cal R}})$ are given as 
\bsub\label{4-15}
\beq
& &{\rm (i)}\ \ {\hat {\cal R}}=\pm\frac{1}{2}\left({\hat T}_m-{\hat T}+p{\hat Q}_p \mp 1\right)\ , \qquad
F_p({\hat {\cal R}})={\hat {\cal R}}\left({\hat {\cal R}}+1\right)\ , 
\label{4-15a}\\
& &{\rm (ii)}\ {\hat {\cal R}}=\pm\frac{1}{2}\left({\hat T}_m-{\hat T}+p{\hat Q}_p \pm 1\right)\ , \qquad
F_p({\hat {\cal R}})={\hat {\cal R}}\left({\hat {\cal R}}-1\right)\ , 
\label{4-15b}
\eeq
\esub
Under the condition (\ref{4-9}), the upper sign of ${\hat {\cal R}}$ in (i) is reduced to ${\hat {\cal S}}$:
\beq\label{4-16}
{\rm If}\quad {\hat Q}_1=1\ , \qquad 
{\hat {\cal R}}={\hat {\cal S}}\quad {\rm and}\quad 
{\hat {\mib {\cal R}}}^2=F_1({\hat {\cal R}})={\hat {\mib {\cal S}}}^2={\hat {\cal S}}\left({\hat {\cal S}}+1\right)\ .
\eeq
The above argument suggests that ${\hat {\cal R}}$ shown in the relation (\ref{4-15}) can be regarded as the 
operator which plays the same role as that of ${\hat {\cal S}}$. 
Then, it may be permitted to require the condition 
\beq\label{4-17}
r_t \geq 0\ . \qquad \left({\hat {\cal R}}\ket{t,t_0}=r_t\ket{t,t_0}\right)
\eeq
The above is formal aspect of the case $p=\pm 1$. 
Later, we will discuss some concrete examples.

Next, we discuss the case $p=0$. 
Under the condition (\ref{4-5}), the relation (\ref{4-10}) is reduced to 
\bsub\label{4-18}
\beq
& &{\hat {\cal R}}_+{\hat {\cal R}}_-={\hat T}_m-{\hat T}+1-\left({\hat T}_m-{\hat T}+1\right)
\sum_t\ket{t}\bra{t}\ , 
\label{4-18a}\\
& &{\hat {\cal R}}_-{\hat {\cal R}}_+={\hat T}_m-{\hat T}\ . 
\label{4-18b}
\eeq
\esub
In this case, ${\hat {\cal R}}_0$ cannot be defined, because of the commutation relation 
\beq\label{4-19}
[\ {\hat {\cal R}}_+\ , \ {\hat {\cal R}}_-\ ]=1-\left({\hat T}_m-{\hat T}+1\right)\sum_t\ket{t}\bra{t}\ . 
\eeq
But, it should be noticed that ${\hat T}_0$ plays the same role as that of ${\hat {\cal R}}_0$ in the case $p=\pm 1$: 
\beq\label{4-20}
[\ {\hat T}_0\ , \ {\hat {\cal R}}_{\pm}\ ]=\pm {\hat {\cal R}}_{\pm}\ . 
\eeq
Of course, ${\hat {\mib {\cal R}}}^2$ cannot be defined. 
The operators ${\hat {\cal R}}_{\pm}$ can be expressed as 
\bsub\label{4-21}\beq
& &{\hat {\cal R}}_+={\hat T}_+\cdot \sqrt{{\hat T}_m-{\hat T}_0}\cdot\left(\sqrt{{\hat T}_-{\hat T}_++\epsilon}\right)^{-1}\ , 
\label{4-21a}\\
& &{\hat {\cal R}}_-=\left(\sqrt{{\hat T}_-{\hat T}_++\epsilon}\right)^{-1}\cdot\sqrt{{\hat T}_m-{\hat T}_0}\cdot {\hat T}_-\ . 
\label{4-21b}
\eeq
\esub
Although ${\hat {\cal R}}_{\pm}$ do not form any algebra, ${\hat {\cal R}}_+$ plays a role of the 
raising operator for constructing the orthogonal set. 
It is easily seen that there exist the relations ${\hat {\cal R}}_-\ket{t,t}=0$ and 
${\hat {\cal R}}_+\ket{t,t_m}=0$ and the state 
$\ket{t,t_0}$ is of the form $({\hat {\cal R}}_+)^{t_0-t}\ket{t}$.

As was already promised, we show concrete examples for the case $p=\pm 1$ and examine the following cases: 
\bsub\label{4-22}
\beq
& &{\rm (i)} \ \ \ {\hat {\cal R}}_0\ket{t,t}=-{\hat {\cal R}}\ket{t,t}\ (=-r_t\ket{t,t})\ , 
\label{4-22a}\\
& &{\rm (ii)} \ \ {\hat {\cal R}}_0\ket{t,t_m}={\hat {\cal R}}\ket{t,t_m}\ (=r_t\ket{t,t_m})\ , 
\label{4-22b}\\
& &{\rm (iii)}\ {\hat {\cal R}}_0\ket{t,t}={\hat {\cal R}}\ket{t,t}\ (=r_t\ket{t,t})\ . 
\label{4-22c}
\eeq
\esub
These three may be regarded as the cases in which traces of the original $su(2)$- and $su(1,1)$-algebras 
are left. 
After rather lengthy consideration, the following results are obtained: 
In the case (i), ${\hat Q}_1=1$, i.e., $q_{1,t}=1$, which leads us to 
\bsub\label{4-23}
\beq
{\hat {\cal R}}=\frac{1}{2}\left({\hat T}_m-{\hat T}\right)\ , \qquad
{\hat {\cal R}}_0={\hat T}_0-\frac{1}{2}\left({\hat T}_m+{\hat T}\right)\ , \qquad 
{\hat {\mib {\cal R}}}^2={\hat {\cal R}}\left({\hat {\cal R}}+1\right)\ . 
\label{4-23a}
\eeq
In the case (ii), $q_{1,t}>0$, which gives 
\beq
& &{\hat {\cal R}}=\frac{1}{2}\left({\hat T}_m-{\hat T}+{\hat Q}_1-1\right)\ , \qquad
{\hat {\cal R}}_0={\hat T}_0-\frac{1}{2}\left({\hat T}_m+{\hat T}-{\hat Q}_1+1\right)\ , 
\nonumber\\
& &{\hat {\mib {\cal R}}}^2={\hat {\cal R}}\left({\hat {\cal R}}+1\right)-\frac{1}{2}
\left({\hat T}_m-{\hat T}+1\right)\left({\hat Q}_1-1\right)\sum_t\ket{t}\bra{t}\ . 
\label{4-23b}
\eeq
\esub
In the case (iii), it is impossible to find any case which satisfies the conditions (\ref{4-8}) and (\ref{4-17}).

It is important to see that the case (i) is included in the case (ii). 
If ${\hat Q}_1=1$ in the case (ii), it is nothing but the case (i) and this case corresponds to the $su(2)$-algebra already discussed. 
If ${\hat Q}_1\neq 1$, there does not exist the relation ${\hat {\cal R}}_0\ket{t,t}=-{\hat {\cal R}}\ket{t,t}$. 
For example, if ${\hat Q}_1=2{\hat T}$, the operator ${\hat Q}_p+p({\hat T}_0-{\hat T})$ for $p=1$ becomes $({\hat T}+{\hat T}_0)$. 
Then, in this case, ${\hat {\cal R}}_{\pm,0}$ can be expressed in the form 
\bsub\label{4-24}
\beq
& &{\hat {\cal R}}_+={\hat T}_+\cdot\sqrt{{\hat T}_m-{\hat T}_0}\cdot\left(\sqrt{{\hat T}_0-{\hat T}+1+\epsilon}\right)^{-1}\ , 
\qquad\qquad\qquad\qquad\qquad\qquad\qquad\qquad
\label{4-24a}\\
& &{\hat {\cal R}}_-=\left(\sqrt{{\hat T}_0-{\hat T}+1+\epsilon}\right)^{-1}\cdot\sqrt{{\hat T}_m-{\hat T}_0}\cdot{\hat T}_-\ , 
\label{4-24b}\\
& &{\hat {\cal R}}_0={\hat T}_0-\frac{1}{2}\left({\hat T}_m-{\hat T}+1\right)\ . 
\label{4-24c}
\eeq
\esub
\beq\label{4-25}
\ \ \ \ \ 
{\hat {\cal R}}=\frac{1}{2}\left({\hat T}_m-{\hat T}-1\right), \ \ 
{\hat {\mib {\cal R}}}^2={\hat {\cal R}}\left({\hat {\cal R}}+1\right)-\frac{1}{2}
\left({\hat T}_m-{\hat T}+1\right)\left(2{\hat T}-1\right)\sum_t\ket{t}\bra{t}. 
\nonumber\\
& &
\eeq
Of course, the following relations are obtained: 
\beq
& &[\ {\hat {\cal R}}_+\ , \ {\hat {\cal R}}_-\ ]=2{\hat {\cal R}}_0-\left({\hat T}_m-{\hat T}+1\right)\left(2{\hat T}-1\right)\sum_t\ket{t}\bra{t}\ , 
\label{4-26}\\
& &{\hat {\cal R}}_0\ket{t,t}=-\frac{1}{2}(t_m-3t+1)\ket{t,t}\neq -{\hat {\cal R}}\ket{t,t}\ . \qquad
({\rm if}\ \ t>1/2)
\label{4-27}
\eeq
The case $t=1/2$ is reduced to the $su(2)$-algebra $({\hat {\cal S}}_{\pm,0})$. 
The above argument may be permitted to call ${\hat Q}_1\neq 1$ as a pseudo $su(2)$-algebra.

The $(t_0-t)$-time operation of ${\hat {\cal R}}_+$ on the state $\ket{t}$ is given in the form 
\beq\label{4-28}
\left({\hat {\cal R}}_+\right)^{t_0-t}\ket{t}
=\sqrt{\frac{(t_m-t)!}{(t_m-t_0)!}}\sqrt{\prod_{k=0}^{t_m-t-1}(q_{p,t}+pk)}\sqrt{\frac{(2t-1)!}{(t_0-t)!(t_0+t-1)!}}\left({\hat T}_+\right)^{t_0-t}\ket{t}\ . 
\eeq
Here, $\prod_{k=0}^{t_m-t-1}(q_{p,t}+pk)$ can be expressed in terms of the gamma-function: 
\beq\label{4-29}
\prod_{k=0}^{t_m-t-1}(q_{p,t}+pk)=p^n\frac{\Gamma(q_{p,t}/p+n)}{\Gamma(q_{p,t}/p)}\ .\qquad (n=t_0-t)
\eeq
If we intend to describe the system under investigation exactly, 
it may be enough to use the orthogonal state $({\hat T}_+)^{n}\ket{t}$ $(n=0,1,2,\cdots ,t_m-t)$. 
However, if we adopt approximation such as in (A), the above 
idea of the deformation becomes useful. 
We will consider the case on the following state: 
\beq\label{4-30}
\ket{\phi_{p,t}}=\frac{1}{\sqrt{\Gamma_{p,t}}}\exp \left(z{\hat {\cal R}}_+\right)\ket{t}\ . \qquad
(\bra{\phi_{p,t}}\phi_{p,t}\rangle=1)
\eeq
Here, $z$ and $\Gamma_{p,t}$ denote a complex parameter and the normalization constant, respectively. 
In such problem, it is indispensable which form is chosen for 
${\hat {\cal R}}_+$. 
In (A), as ${\hat {\cal R}}_+$, ${\hat {\cal T}}_+$, i.e., ${\hat T}_+$ itself was used. 
The norm of the state $({\hat {\cal R}}_+)^{n}\ket{t}$ is given by 
\beq
\bra{t}\left({\hat {\cal R}}_-\right)^n\!\!\cdot\!\left({\hat {\cal R}}_+\right)^n\ket{t}
\!\!&=&\!\!
\frac{(t_m-t)!}{(t_m-t-n)!}\prod_{k=0}^{n-1}(q_{p,t}+pk)\cdot\frac{(2t-1)!}{n!(2t-1+n)!}\bra{t}\left({\hat T}_-\right)^n\!\!
\cdot\!\left({\hat T}_+\right)^n\ket{t}\nonumber\\
&=&\frac{(t_m-t)!}{(t_m-t-n)!}\prod_{k=0}^{n-1}(q_{p,t}+pk)\ , 
\label{4-31}\\
\bra{t}\left({\hat T}_-\right)^n\!\!\cdot\!\left({\hat T}_+\right)^n\ket{t}
&=&\frac{(2t-1+n)!n!}{(2t-1)!}\ . 
\label{4-32}
\eeq
%
%%%%%%%%%%%%%%%%%%%%%%%%%%%%%%%%%%%%%%%%%%%%%%%%%%%%%%%%%%%%%%%%%%%%%%%%%%%%%%%%%%%%%%%%%%%%%
%
Then, $\Gamma_{p,t}$ is expressed as a function of $x$ in the following: 
\beq
\Gamma_{p,t}\!\!&=&\!\!
\sum_{n=0}^{t_m-t}(-)^n\left(
\begin{array}{c}
t_m-t \\ n
\end{array}
\right)
\frac{\Gamma(q_{p,t}/p+n)}{n!\Gamma(q_{p,t}/p)}(-px)^n \nonumber\\
&=&G_{t_m-t}(q_{p,t}/p-(t_m-t),1;-px)\ , 
\label{4-33}\\
x&=&|z|^2\ . 
\label{4-34}
\eeq
Here, the relation (\ref{4-29}) was used. 
The function $G_{t_m-t}$ is Jacobi polynomial. 
At the limit $p\rightarrow 0$ and $q_{p,t}\rightarrow 1$ for the expression (\ref{4-33}) is reduced to 
\beq\label{4-35}
\Gamma_{0,t}=\lim_{\tiny
\begin{array}{c}
p\rightarrow 0\\
%,\ 
q_{p,t}\rightarrow 1
\end{array}
}\Gamma_{p,t}
=\sum_{n=0}^{t_m-t}\frac{(-)^n}{n!}
\left(
\begin{array}{c}
t_m-t \\ n 
\end{array}
\right)(-x)^n
=L_{t_m-t}(-x)\ . 
\eeq
The function $L_{t_m-t}$ denotes Laguerre polynomial. 
Here, the following relation was used:
\beq\label{4-36}
\lim_{
\tiny
\begin{array}{c}
p\rightarrow 0\\
q_{p,t}\rightarrow 1
\end{array}}
\frac{\Gamma(q_{p,t}/p+n)}{\Gamma(q_{p,t}/p)}(-px)^n=(-x)^n\ . 
\eeq
With the use of the relation (\ref{4-29}), we can obtain the expression (\ref{4-35}) directly.

The expression (\ref{4-33}) is too complicated to apply it to any concrete problem. 
This indicates that idea for the approximation must be searched. 
For this aim, three points for $\Gamma_{p,t}$ must be pointed out. 
First is the case $q_{1,t}=1$, which leads us to the simple form: 
\beq\label{4-37}
\Gamma_{1,t}=(1+x)^{t_m-t}\ .
\eeq
Second is the maximum power of the polynomial $\Gamma_{p,t}$ for $x$:
\beq\label{4-38}
{\rm the\ maximum\ power}=t_m-t\ .
\eeq
It is independent of the choice of $q_{p,t}$. 
Third is related to the behavior of $\Gamma_{p,t}$ near $x=0$. 
In the region $x\sim 0$, $\Gamma_{p,t}$ can be expressed as 
\bsub\label{4-39}
\beq
& &\Gamma_{p,t}=1+\Gamma_{p,t}^{(1)} x+\frac{1}{2}\Gamma_{p,t}^{(2)}x^2+\cdots\ , 
\label{4-39a}\\
& &\Gamma_{p,t}^{(1)}=(t_m-t)q_{p,t}\ , \qquad
\Gamma_{p,t}^{(2)}=\frac{1}{2}(t_m-t)(t_m-t-1)q_{p,t}(q_{p,t}+p) .
\label{4-39b}
\eeq
\esub
Concerning $\Gamma_{p,t}(x)$, the above-mentioned three points suggest us the following approximation: 
\beq\label{4-40}
& &\Gamma^a(x)=(1+Cx)^k(1+Dx)^l\ , \nonumber\\
& &k\ , \ \ l\ :\ {\rm positive\ integers}\ , \quad k+l=m\ (=t_m-t)\ . 
\eeq
In order to take into account the difference between $C$ and $D$, 
$\Gamma^a(x)$ should be treated in the region 
\beq\label{4-41}
1 \leq l \leq m-1\ .
\eeq
Hereafter, in order to avoid unnecessary complication, we will omit 
the index $(p,t)$ and use the symbol $m(=t_m-t)$. 
We determine $C$ and $D$ so as to make the coefficients of the terms $x$ and $x^2$ in the 
form (\ref{4-40}) agree with those in the relation (\ref{4-39}):
\beq\label{4-42}
C=q\left(1\mp \sqrt{\frac{l}{k}\cdot\frac{m-1}{2}\cdot \zeta}\right)\ , \qquad
D=q\left(1\pm\sqrt{\frac{k}{l}\cdot\frac{m-1}{2}\cdot \zeta}\right)\ , 
\eeq
\vspace{-0.2cm}
\bsub\label{4-43}
\beq\label{4-43a}
\zeta=1-\frac{p}{q}\ (\geq 0)\ . 
\qquad\qquad\qquad\qquad\qquad\qquad\qquad\qquad\qquad\qquad
\eeq
The condition $\zeta\geq 0$ and the relations (\ref{4-5}) and (\ref{4-8}) give us the inequality for $\zeta$:
\beq
& &0\leq \zeta \leq 1+\frac{1}{m}\ , \nonumber\\
{\rm i.e.,}& &\nonumber\\
& &0\leq \zeta <1\ (p=1)\ , \qquad \zeta=1\ (p=0)\ , \qquad 
1< \zeta \leq 1+\frac{1}{m}\ (p=-1)\ .\qquad
\label{4-43b}
\eeq
\esub
The function $\Gamma(x)(=\Gamma_{p,t}(x))$ is a polynomial, in which the coefficient of 
each term is positive and, then, $\Gamma(x)=0$ does not have any root 
in the range $x\geq 0$, i.e., $\Gamma(x)>0\ (x\geq 0)$. 
Therefore, we require the condition $\Gamma^a(x)>0\ (x\geq 0)$, that is, 
\beq\label{4-44}
C>0\ , \qquad D>0\ . 
\eeq
The quantities $C$ and $D$ are symmetric with each other for $k$, $l$, $\mp$ and $\pm$ 
in the relation (\ref{4-42}) and for $D$, we can adopt the form with the upper in the 
sign $\pm$: 
\bsub\label{4-45}
\beq
D=q\left(1+\sqrt{\frac{k}{l}\cdot\frac{m-1}{2}\cdot \zeta}\right)\ (>0)\ . 
\label{4-45a}
\eeq
Therefore, automatically, $C$ becomes of the form 
\beq
C=q\left(1-\sqrt{\frac{l}{k}\cdot\frac{m-1}{2}\cdot \zeta}\right)\ (>0)\ . 
\label{4-45b}
\eeq
\esub
The expression (\ref{4-45a}) satisfies $D>0$. 
The condition $C>0$ is reduced to 
\beq\label{4-46}
l < \lambda_0\ , \qquad
\lambda_0=\frac{m}{1+\frac{m-1}{2}\cdot \zeta}\ . 
\eeq
%
%%%%%%%%%%%%%%%%%%%%%%%%%%%%%%%%%%%%%%%%%%%%%%%%%%%%%%%%%%%%%%%%%%%
\begin{table}[t]
\caption{Parameter set}%%%Table caption goes here
\label{table1}
\centering
\begin{tabular}{c|c|c|c}%%%The number of columns has to be defined here
\hline
$l_0$ & $\zeta_{\rm min}$ & $\zeta_{\rm max}$ &$l$\\ 
\hline
$m-1$ & 0 & $\frac{2}{(m-1)^2}$ & $m-1,\ m-2,\ \cdots ,\ 2,\ 1$\\
$m-2$ & $\frac{2}{(m-1)^2}$ & $\frac{4}{(m-1)(m-2)}$ & $m-2,\ m-3,\ \cdots ,\ 2,\ 1$\\
$\vdots$ & $\vdots$ & $\vdots$ & $\vdots$\\
$\frac{m+1}{2}$ & $\frac{2(m-3)}{(m-1)(m+3)}$ & $\frac{2}{m+1}$ & $\frac{m+1}{2},\ \frac{m-1}{2},\ \cdots ,\ 2,\ 1$\ 
($m$ : odd)\\
$\frac{m}{2}$ & $\frac{2(m-2)}{(m-1)(m+2)}$ & $\frac{2}{m-1}$ & $\frac{m}{2},\ \frac{m-2}{2},\ \cdots ,\ 2,\ 1$\ 
($m$ : even)\\
$\frac{m-1}{2}$ & $\frac{2}{m+1}$ & $\frac{2(m+1)}{(m-1)^2}$ & $\frac{m-1}{2},\ \frac{m-3}{2},\ \cdots ,\ 2,\ 1$\ 
($m$ : odd)\\
$\vdots$ & $\vdots$ & $\vdots$ & $\vdots$\\
$2$ & $\frac{2(m-3)}{3(m-1)}$ & $\frac{m-2}{m-1}$ & 2,\ 1\\
$1$ & $\frac{m-2}{m-1}$ & $1+\frac{1}{m}$ & 1\\
%%%% Table body
\hline
%Parameter set II & 0.800 & 20.0 & 6.6 \\
%\hline
\end{tabular}
\end{table}%%%End of the table
%%%%%%%%%%%%%%%%%%%%%%%%%%%%%%%%%%%%%%%%%%%%%%%%%%%%%%%%%%%%%%%%%%%%%%%%%%%%%%%%%%%%%%%%
%
It should be noted that $l$ is a positive integer, but, $\lambda_0$ is generally not an integer. 
We introduce $l_0$ which is the nearest integer to $\lambda_0$ under the 
condition $l_0<\lambda_0$. 
Then, the following inequality is obtained: 
\beq\label{4-47}
0 < \lambda_0-l_0 \leq 1\ , \qquad
l=l_0,\ l_0-1,\cdots ,\ 2,\ 1.
\eeq
With the use of the relation (\ref{4-46}), the inequality (\ref{4-47}) can be rewritten in the form 
\beq\label{4-48}
& &\zeta_{\rm min} \leq \zeta \leq \zeta_{\rm max}\ , \nonumber\\
& &\zeta_{\rm min}=\frac{2(m-1-l_0)}{(m-1)(l_0+1)}\ , \qquad
\zeta_{\rm max}=\frac{2(m-l_0)}{(m-1)l_0}\ . 
\eeq
Some concrete cases of the inequality are summarized in Table I. 
We can see that the case $(p=1,\ q=1)$ reduces to $\zeta=0$ and $C=D=1$. 
In this case, the relation (\ref{4-40}) becomes to the form (\ref{4-37}) independent of the 
choice of $l$. 
Of course, this case corresponds to the $su(2)$-algebra. 
The result in any other case depends on the choice of $l$. 
Up to the present, we have no idea how to determine the value of $l$. 
In relation to a certain approximation adopted in next section, the case $l=1$ may be the most reasonable. 

%%%%%%%%%%%%%%%%%%%%%%%%%%%%%%%%%%%%%%%%%%%%%%%%%%%%%%%%%%%%%%%%%%%%%%
\begin{figure}[b]
\begin{center}
\includegraphics[height=5.0cm]{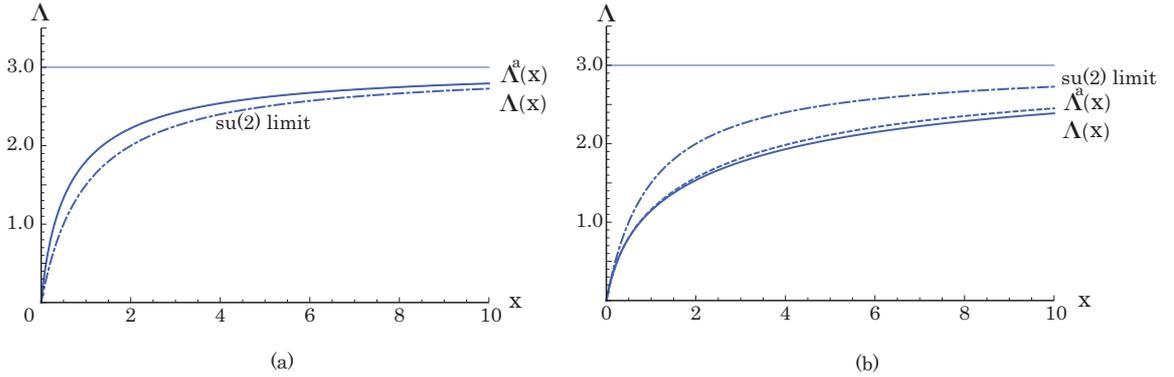}
\caption{
Figure (a) shows the comparison of $\Lambda^a(x)$ with $\Lambda (x)$ in Eq. (\ref{4-52}) for the case $k=m-1,\ l=1$ with $m=3$. 
Here, $p=1$ and $q=2$ are adopted. In this case, $\Lambda^a(x)$ and $\Lambda(x)$ almost overlap one another. 
Figure (b) shows the comparison of $\Lambda^a(x)$ (dotted curve) with $\Lambda (x)$ (solid curve) in Eq. (\ref{4-53}) with $p=0$ and $q=1$. 
On both the figures, the dash-dotted curves represent the $su(2)$ limit. }
\label{fig:1}
\end{center}
\end{figure}
%%%%%%%%%%%%%%%%%%%%%%%%%%%%%%%%%%%%%%%%%%%%%%%%%%%%%%%%%%%%%%%%%%%%%%%%

The expectation values of ${\hat T}_0$, ${\hat {\cal R}}_+$ and ${\hat {\cal R}}_-$ for $\ket{\phi}$ given 
in the relation (\ref{4-30}) are calculated in the form 
\bsub\label{4-49}
\beq
& &\bra{\phi}{\hat T}_0\ket{\phi}=t+\Lambda(x)\ , 
\label{4-49a}\\
& &\bra{\phi}{\hat {\cal R}}_+\ket{\phi}=z^*\frac{\Lambda(x)}{x}\ , \qquad
\bra{\phi}{\hat {\cal R}}_-\ket{\phi}=z\frac{\Lambda(x)}{x}\ . 
\label{4-49b}
\eeq
\esub
Here, $\Lambda(x)$ is defined as 
\beq\label{4-50}
\Lambda(x)=\frac{x\frac{d\Gamma(x)}{dx}}{\Gamma(x)}\ . 
\eeq
In the case of the approximate form of $\Lambda(x)$, $\Lambda^a(x)$ is given in the form 
\beq\label{4-51}
\Lambda^a(x)=\frac{kCx}{1+Cx}+\frac{lDx}{1+Dx}
=m\left[1-\left(\frac{k/m}{1+Cx}+\frac{l/m}{1+Dx}\right)\right]\ . 
\eeq
The relation (\ref{4-51}) will play a central role in next section. 
The relation (\ref{4-34}) and (\ref{4-35}) lead us to the following expression for $\Lambda(x)$: 
\beq
& &\Lambda(x)=m\left(1-\frac{G_{m-1}(q/p-m+1,1,-px)}{G_m(q/p-m,1,-px)}\right)\ , 
\label{4-52}\\
& &\Lambda(x)=m\left(1-\frac{L_{m-1}(-x)}{L_m(-x)}\right)\ . 
\label{4-53}
\eeq
Figure \ref{fig:1} shows the comparison of $\Lambda^a(x)$ for the case $(k=m-1,\ l=1)$ and 
$\Lambda(x)$.

\setcounter{equation}{0}

\section{Application}

As was mentioned in \S 1, the aim of this paper is to formulate a new boson representation of the $su(2)$-algebra and its deformation, 
in which the idea of the phase space doubling is applied straightforwardly. 
In this section, in order to demonstrate our idea, 
we will apply the present form to the case of a simple boson model. 
This model is essentially the same as that discussed in \S 7 in (A). 
We pay an attention to the boson Hamiltonian 
\beq\label{5-1}
{\hat H}_b=\omega {\hat b}^*{\hat b}\ . 
\eeq
The Hamiltonian ${\hat H}_b$ is nothing but ${\hat H}_{\rm intr}$ introduced in the 
relation (\ref{1-1}). 
Following the idea of the phase space doubling, we introduce another boson Hamiltonian ${\hat H}_a=\omega{\hat a}^*{\hat a}$ and set 
up the form 
\beq\label{5-2}
{\hat H}_{ba}={\hat H}_b-{\hat H}_a=\omega({\hat b}^*{\hat b}-{\hat a}^*{\hat a})\ . 
\eeq
Of course, ${\hat H}_a$ plays a role of ${\hat H}_{\rm extr}$ and ${\hat H}_{ba}={\hat H}_0$. 
As for the interaction between two boson systems, 
${\hat V}_{ba}$, we adopt the following form: 
\beq\label{5-3}
{\hat V}_{ba}=-i\gamma({\hat a}^*{\hat b}^*\cdot f({\hat a}^*{\hat a},\ {\hat b}^*{\hat b})
-f({\hat a}^*{\hat a},\ {\hat b}^*{\hat b})\cdot {\hat b}{\hat a})\ .
\eeq
Here, $\gamma$ denotes the interaction strength. 
For example, the case $f({\hat a}^*{\hat a},\ {\hat b}^*{\hat b})=1$ corresponds to the $su(1,1)$-algebraic model investigated in Refs.\citen{5,6,7}.
As for ${\hat a}^*{\hat b}^*\cdot f({\hat a}^*{\hat a},\ {\hat b}^*{\hat b})$, we adopt the operator ${\hat {\cal R}}_+$ shown in the 
relation (\ref{4-3}) and, then, the Hamiltonian ${\hat H}$ is given by 
\beq\label{5-4}
{\hat H}={\hat H}_{ba}+{\hat V}_{ba}
=\omega\left(2{\hat T}-1\right)-i\gamma\left({\hat {\cal R}}_+-{\hat {\cal R}}_-\right)\ . 
\eeq
It should be noted that ${\hat H}$ does not mean the total energy. 
It may be clear that ${\hat T}$ is a constant of motion. 
The above is our model discussed in this paper.

For treating the Hamiltonian (\ref{5-4}), we follow the same method as that in (A). 
Regarding $\ket{\phi}$ as a time-dependent variational state, we set up the following variational equation: 
\beq\label{5-5}
\delta \int \bra{\phi}i\partial_{\tau}-{\hat H}\ket{\phi}d\tau =0 \ . 
\eeq
Here, in order to avoid confusion between the time variable and the quantum number $t$, we will use $\tau$ for the time variable. 
If $z$ and $z^*$ are regarded as time-dependent variational parameters, 
the variational equation (\ref{5-5}) leads us to the following equation: 
\beq\label{5-6}
{\dot z}=-\gamma\left[1-\frac{z^2}{x}\left(1-\frac{\Lambda(x)}{x\frac{d\Lambda(x)}{dx}}\right)\right]\ , \quad
{\dot z}^*=-\gamma\left[1-\frac{z^{*2}}{x}\left(1-\frac{\Lambda(x)}{x\frac{d\Lambda(x)}{dx}}\right)\right]\ . 
\eeq
The expectation value of ${\hat H}$, ${{\cal H}}$, is given in the form 
\beq\label{5-7}
{\cal H}&=&\bra{\phi}{\hat H}\ket{\phi}=\omega(2t-1)-i\gamma({\cal R}_+-{\cal R}_-)\nonumber\\
&=&\omega(2t-1)-\gamma i(z^*-z)\frac{\Lambda(x)}{x}\ . 
\eeq
Here, ${\cal R}_{\pm}$ denote the expectation values of ${\hat {\cal R}}_{\pm}$. 
The detail can be found in (A).

The present system is of two dimension and, therefore, there exist two constants of motion. 
One is the quantum number $t$ and the second, which will be denoted as $\kappa$, is given through the relation 
\beq\label{5-8}
i(z^*-z)\frac{\Lambda(x)}{x}=2\kappa\ . 
\eeq
It may be self-evident, because ${\cal H}$ itself shown in the relation (\ref{5-8}) is a 
constant of motion. 
If $z$ is expressed in the form $z=u+iv$, we have 
\beq\label{5-9}
i(z^*-z)=2v\ . 
\eeq
In (A), we learned that, instead of $x$, it may be convenient to adopt the variable $y$ defined as 
\beq\label{5-10}
y=\frac{\Lambda(x)}{x}\ . \qquad (x=|z|^2=u^2+v^2)
\eeq
Inversely solving, $x$ can be expressed as a function of $y$. 
Then, $v$ can be expressed in the form 
\beq\label{5-11}
v=\frac{\kappa}{y}\ , \qquad {\rm i.e.,}\qquad 
yu=\pm\sqrt{xy^2-\kappa^2}\ . 
\eeq
With the use of the relation (\ref{5-6}), ${\dot x}$ can be given as 
\beq\label{5-12}
{\dot x}=-2\gamma\frac{\Lambda(x)}{x\frac{d\Lambda(x)}{dx}}\cdot u\ . 
\eeq
The definitions of $y$ and ${\dot x}$, which are given in the relations (\ref{5-10}) and 
(\ref{5-12}), respectively, give us ${\dot y}$ in the form 
\beq\label{5-13}
{\dot y}=-\frac{2\gamma}{x+y\frac{dx}{dy}}\cdot\left(\pm\sqrt{xy^2-\kappa^2}\right)\ . 
\eeq

Now, let us express $x$ as a function of $y$. 
Basic equation of this task is the relation (\ref{5-10}). 
As for $\Lambda(x)$, we adopt the approximate form $\Lambda^a(x)$ given in the relation (\ref{4-51}). 
For $\Lambda^a(x)$, the relation (\ref{5-10}) is reduced to the form 

\beq\label{5-14}
CDy\cdot x^2-\left(mCD-y(C+D)\right)\cdot x+\left(y-(kC+lD)\right)=0\ . 
\eeq
A solution of Eq.(\ref{5-14}) is as follows: 
\beq\label{5-15}
& &x=\frac{m}{2y}-\frac{C+D}{2CD}+\frac{m}{2y}\sqrt{1+2Iy+J^2y^2}\ , \nonumber\\
& &I=\left(\frac{k-l}{m^2}\right)\left(\frac{C-D}{CD}\right)\ , \qquad
J^2=\frac{1}{m^2}\left(\frac{C-D}{CD}\right)^2\ . \qquad
\left(J^2=\left(\frac{k+l}{k-l}\right)^2I^2\right)\qquad
\eeq
In the case $C=D$, another solution becomes negative and we pick up only the solution (\ref{5-15}). 
Next, we consider a possible approximation of $\sqrt{1+2Iy+J^2y^2}$, which, up to the term $y^2$, is expanded for $y$: 
\beq\label{5-16}
\sqrt{1+2Iy+J^2y^2}=1+Iy+\frac{1}{2}(J^2-I^2)y^2\ . 
\eeq
Let the following inequality be permitted: 
\beq\label{5-17}
y \ll \left|\frac{2I}{J^2-I^2}\right|\ . 
\eeq
Then, we are able to obtain the approximate form 
\beq\label{5-18}
\sqrt{1+2Iy+J^2y^2}=1+Iy=1+\left(\frac{k-l}{m^2}\right)\left(\frac{C-D}{CD}\right)y\ . 
\eeq
Later, we will discuss the condition, under which the inequality (\ref{5-17}) is meaningful. 
Then, we have 
\beq
& &x=\frac{m}{2y}-\frac{C+D}{2CD}+\frac{m}{2y}\left(1+\left(\frac{k-l}{m^2}\right)\left(\frac{C-D}{CD}\right)y\right)
=\frac{m}{y}-\frac{1}{B}\ , 
\label{5-19}\\
& &\frac{1}{B}=\frac{1}{m}\left(\frac{k}{C}+\frac{l}{D}\right)\ . 
\label{5-20}
\eeq

With the use of the relation (\ref{5-19}), we obtain 
\bsub\label{5-21}
\beq
& &xy^2-\kappa^2=\left(\frac{m^2B}{4}-\kappa^2\right)-\frac{1}{B}\left(y-\frac{mB}{2}\right)^2\ , 
\label{5-21a}\\
& &x+y\frac{dx}{dy}=-\frac{1}{B}\ . 
\label{5-21b}
\eeq
\esub
Therefore, ${\dot y}$ shown in the relation (\ref{5-13}) can be expressed as 
\beq\label{5-22}
{\dot y}=\pm 2\gamma B\sqrt{\left(\frac{m^2B}{4}-\kappa^2\right)-\frac{1}{B}\left(y-\frac{mB}{2}\right)^2}\ . 
\eeq
By solving Eq.(\ref{5-22}), $y$ can be expressed as a function $\tau$. 
The relation (\ref{5-22}) can be rewritten to the form 
\beq\label{5-23}
\frac{1}{2}{\dot y}^2+\frac{1}{2}\cdot 4\gamma^2 B\left(y-\frac{mB}{2}\right)^2
=2\gamma^2 B\left[\left(\frac{mB}{2}\right)^2-\left(\sqrt{B}\kappa\right)^2\right]\ . 
\eeq
The relation (\ref{5-23}) tells us that the present system is equivalent to a simple harmonic oscillator in the classical mechanics. 
Then, we have 
\beq\label{5-24}
y=\frac{mB}{2}+\sqrt{\left(\frac{mB}{2}\right)^2-\left(\sqrt{B}\kappa\right)^2}\cos(2\gamma\sqrt{B}\tau+\chi_0)\ . 
\eeq
Here, $\chi_0$ is determined by the initial condition. 
The quantities $x$ and $\Lambda^a(x)$ can be expressed in the following form: 
\beq
& &x=\frac{\frac{mB}{2}-\sqrt{\left(\frac{mB}{2}\right)^2-\left(\sqrt{B}\kappa\right)^2}\cos(2\gamma\sqrt{B}\tau+\chi_0)}
{\frac{mB}{2}+\sqrt{\left(\frac{mB}{2}\right)^2-\left(\sqrt{B}\kappa\right)^2}\cos(2\gamma\sqrt{B}\tau+\chi_0)}\ , 
\label{5-25}\\
& &\Lambda^a(x)=\frac{mB}{2}-\sqrt{\left(\frac{mB}{2}\right)^2-\left(\sqrt{B}\kappa\right)^2}\cos(2\gamma\sqrt{B}\tau+\chi_0)\ . 
\label{5-26}
\eeq
Thus, we could express $x$ and $\Lambda^a(x)$ as functions of $\tau$. 
Of course, it is a general solution, in which the initial and the boundary condition are not taken into account. 
In \S 6, we will discuss these conditions. 
%%
%
%

%%%%%%%%%%%%%%%%%%%%%%%%%%%%%%%%%%%%%%%%%%%%%%%%%%%%%%%%%%%%%%%%%%%%%%
\begin{figure}[t]
\begin{center}
\includegraphics[height=6.5cm]{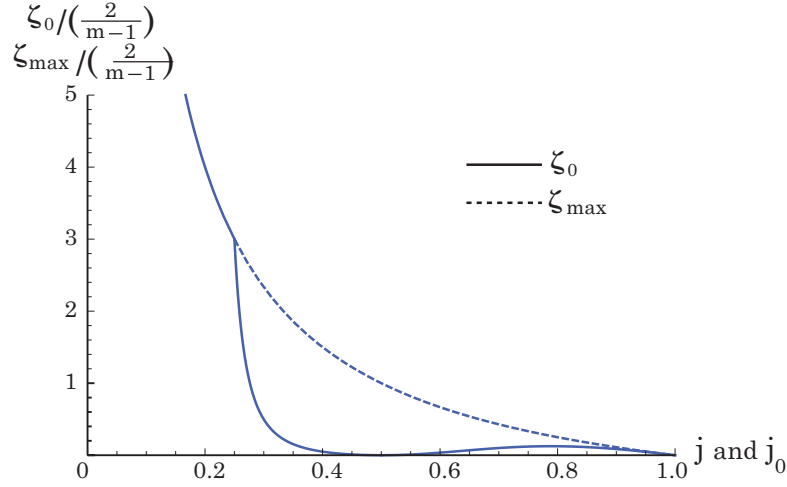}
\caption{For the comparison, $\zeta_0(j)$ in Eq.(\ref{5-30}) and $\zeta_{\rm max}$ in Eq.(\ref{5-32}) are depicted. 
The vertical axis represents $\zeta_0(j)$ and/or $\zeta_{\rm max}(j_0)$ in unit $2/(m-1)$. 
}
\label{fig:2}
\end{center}
\end{figure}
%%%%%%%%%%%%%%%%%%%%%%%%%%%%%%%%%%%%%%%%%%%%%%%%%%%%%%%%%%%%%%%%%%%%%%%%

Finally, we will discuss the inequality (\ref{5-17}), 
which leads us to the simple result shown in the relation (\ref{5-26}). 
First, we note the maximum value of $y$, $y_{\rm max}$, which is expressed as 
\beq\label{5-27}
y_{\rm max}=mB\ . 
\eeq
The relation (\ref{5-27}) is obtained under the condition $(\kappa=0,\ \cos(2\gamma\sqrt{B}\tau+\chi_0)=1)$ 
in the relation (\ref{5-24}). 
Therefore, we have 
\beq\label{5-28}
mB\ll \left|\frac{2I}{J^2-I^2}\right|\ . 
\eeq
After rather lengthy consideration, the following inequality can be derived from the relation (\ref{5-28}): 
\beq\label{5-29}
\zeta_0 \ll \zeta_0(j)\ , 
\qquad\qquad\qquad\qquad\qquad
\eeq
\bsub\label{5-30}
\beq
& &{\rm (i)\ \ \ for}\ \ 0<j\leq \frac{1}{4}\ , \quad \zeta_0(j)=\frac{2}{m-1}\cdot\frac{1-j}{j}\ , 
\label{5-30a}\\
& &{\rm (ii)\ \ for}\ \ \frac{1}{4}<j < \frac{1}{2}\ , \quad \zeta_0(j)=\frac{2}{m-1}\cdot\frac{j(1-j)(1-2j)^2}{(1-6j+6j^2)^2}\ , 
\label{5-30b}\\
& &{\rm (iii)\ for}\ \ \frac{1}{2}\leq j < 1\ , \quad \zeta_0(j)=\frac{2}{m-1}\cdot\frac{j(1-j)(1-2j)^2}{(1-2j+2j^2)^2}\ . 
\label{5-30c}
\eeq
\esub
Here, $j$ denotes 
\beq\label{5-31}
j=\frac{l}{m}\ . \qquad (0 < j < 1)
\eeq
On the other hand, we note the inequality (\ref{4-48}):
\beq\label{5-32}
\zeta < \zeta_{\rm max}(j_0)\ , \qquad
\zeta_{\rm max}(j_0)=\frac{2}{m-1}\cdot\frac{1-j_0}{j_0}\ . 
\eeq
Here, $j_0$ denotes 
\beq\label{5-33}
j_0=\frac{l_0}{m}\ . \qquad (0 < j_0 < 1)
\eeq
Further, we note the relation (\ref{4-40}), which can be expressed as 
\beq\label{5-34}
j_0 \geq j\ . 
\eeq
The inequality (\ref{5-29}) and (\ref{5-32}) suggest us the relation 
\beq\label{5-35}
\zeta_{\rm max}(j_0) \ll \zeta_0(j)\ . 
\eeq
Figure \ref{fig:2} shows the behavior of $\zeta_0$ and $\zeta_{\rm max}$ in unit $(2/(m-1))$. 
From the figure, we can learn the following points: 
(i) If $j\sim 0$ and $j_0 \sim 1$, the inequality (\ref{5-35}) is sufficiently satisfied. 
(ii) If $1/4 \leq j < 1/2$ and $j_0\sim 1$, the inequality (\ref{5-35}) may be satisfied, but not so sufficient as the case (i). 
(iii) If $1/2 < j < 1$ and $1/2 < j_0 < 1$, the inequality (\ref{5-35}) is not satisfied. 
The above summarize gives us 
the following conclusion: 
If $l$ is rather far from $l_0$ $(l \ll l_0)$, 
our approximation may be justified. 
Therefore, the case $(l=1,\ l_0=m-1)$ is the most reliable. 
This point has been already suggested in the previous section.

\begin{figure}[t]
\begin{center}
\includegraphics[height=5.5cm]{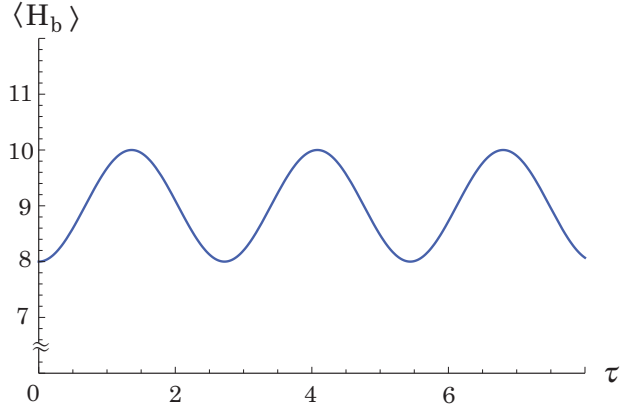}
\caption{The time-dependent energy for $b$-system is depicted as a function of time $\tau$. 
}
\label{fig:3}
\end{center}
\end{figure}

As an example of physical systems, let us consider $b$-system governed by the Hamiltonian (\ref{5-2}) considered in this section. 
Figure \ref{fig:3} shows the energy expectation value for $b$-system as a function of time $\tau$, 
which is depicted by using the approximation in Eq.(\ref{5-26}). 
The parameters are taken as $t=4,\ s=3/2$, which leads to $t_m=7$ and $m=3$. 
Also, $l=1$, $p=1,\, q=2, \gamma=1,\ \omega=1$ and $\kappa=3/2$ are adopted 
and an initial condition, $\chi_0=0$, is given. 
It is seen that the energy flows into $b$-system from external environment and vice versa.
%
%%%%%%%%%%%%%%%%%%%%%%%%%%%%%%%%%%%%%%%%%%%%%%%%%%%%%%%%%%%%%%%%%%%%%%
%%%%%%%%%%%%%%%%%%%%%%%%%%%%%%%%%%%%%%%%%%%%%%%%%%%%%%%%%%%%%%%%%%%%%%%%
%
%
%

\setcounter{equation}{0}
\section{Discussion}

First of all, we will examine the formal result of the approximate 
solution (\ref{5-26}) closely. 
For this aim, first, we consider the quantity $B$ defined in the relation (\ref{5-20}). 
With the use of the relations (\ref{4-45a}) and (\ref{4-45b}), $B$ can be expressed in the form for the case 
($k=m-1,\ l=1$) as follows: 
\beq\label{6-1}
B=q\left(1-\sqrt{\frac{\zeta}{2}}\right)\left(1+\frac{1}{m-\left(2-\sqrt{\frac{2}{\zeta}}\right)}\right)\ . 
\eeq 
Since $m\geq 2$, $B$ obeys the inequality 
\beq\label{6-2}
B\geq q\left(1-\sqrt{\frac{\zeta}{2}}\right)\ . 
\eeq
We are mostly interested in the case $p=1$, i.e., $\zeta=1-1/q$. 
Then, we have the relation 
\beq\label{6-3}
q\left(1-\sqrt{\frac{\zeta}{2}}\right)-1=\frac{1}{\sqrt{2}}\frac{\sqrt{q-1}(q-2)}{\sqrt{2(q-1)}+\sqrt{q}}\geq 0\ . 
\eeq
The inequalities (\ref{6-2}) and (\ref{6-3}) lead us to 
\beq\label{6-4}
B\geq 1\ . 
\eeq
The examples are shown in Table {\ref{table2}}. 
%
%%%%%%%%%%%%%%%%%%%%%%%%%%%%%%%%%%%%%%%%%%%%%%%%%%%%%%%%%%%%%%%%%%%
\begin{table}[t]
\caption{Examples for $q$ and $B$.}%%%Table caption goes here
\label{table2}
\centering
\begin{tabular}{c|l}%%%The number of columns has to be defined here
\hline
$q$ & $B$\\ 
\hline
1 & 1 \\
2 & $1+\frac{1}{m}$ \\
6 & $(6-\sqrt{5})\left(1+\frac{1}{m-2\left(1-\sqrt{\frac{3}{5}}\right)}\right)\approx 2.127\left(1+\frac{1}{m-0.451}\right)$\\
9 & $3\left(1+\frac{1}{m-\frac{1}{2}}\right)$ \\
13 & $(13-\sqrt{78})\left(1+\frac{1}{m-\left(2-\sqrt{\frac{6}{13}}\right)}\right) \approx 4.168\left(1+\frac{1}{m-1.321}\right)$\\
%%%% Table body
\hline
%Parameter set II & 0.800 & 20.0 & 6.6 \\
%\hline
\end{tabular}
\end{table}%%%End of the table
%%%%%%%%%%%%%%%%%%%%%%%%%%%%%%%%%%%%
%
Later at several places, we will use the inequality (\ref{6-4}).

Now, we investigate the general solution (\ref{5-26}). 
For this aim, we define two functions 
\bsub\label{6-5}
\beq
& &\Lambda^a(\theta)=\frac{mB}{2}-\sqrt{\left(\frac{mB}{2}\right)^2-\left(\sqrt{B}\kappa\right)^2}\ \cos\theta\ , 
\label{6-5a}\\
& &y(\theta)=\frac{mB}{2}+\sqrt{\left(\frac{mB}{2}\right)^2-\left(\sqrt{B}\kappa\right)^2}\ \cos\theta\ . 
\label{6-5b}
\eeq
\esub
If $\theta$ is replaced with $(2\gamma\sqrt{B}\tau+\chi_0)$, $\Lambda^a(\theta)$ and $y(\theta)$ are reduced to 
the results (\ref{5-26}) and (\ref{5-24}). 
The present approximate result should obey the following boundary conditions: 
\bsub\label{6-6}
\beq
& &{\rm (i)}\ \ \ 
y(\theta)\cdot\Lambda^a(\theta)-\kappa^2 \geq 0\ , \quad{\rm i.e.,}\quad 
\left(\frac{mB}{2}\right)^2-\left(\left(\frac{mB}{2}\right)^2-\left(\sqrt{B}\kappa\right)^2\right)\cos^2\theta -\kappa^2 \geq 0\ , \nonumber\\
& &
\label{6-6a}\\
& &{\rm (ii)}\ \ 
\left(\frac{mB}{2}\right)^2-\left(\sqrt{B}\kappa\right)^2 \geq 0\ , 
\label{6-6b}\\
& &{\rm (iii)}\ 
\Lambda^a(\theta)\leq m\ , \quad{\rm i.e.,}\quad 
\frac{mB}{2}-\sqrt{\left(\frac{mB}{2}\right)^2-\left(\sqrt{B}\kappa\right)^2}\ \cos\theta \leq m\ . 
\label{6-6c}
\eeq
\esub
The condition (i) results from the relation (\ref{5-11}), in which $(xy^2-\kappa^2)$ 
can be expressed in the form 
$(y(\theta)\cdot\Lambda^a(\theta)-\kappa^2)$. 
The condition (ii) may be self-evident. 
The condition (iii) comes from the relations (\ref{4-52}) and (\ref{4-53}). 
On the basis of the conditions (i), (ii) and (iii), we examine our general solution (\ref{5-26}).

Let us start in the condition (i). 
It is easily verified through the following inequality: 
\beq\label{6-7}
y(\theta)\cdot\Lambda^a(\theta)-\kappa^2=
\left(\left(\frac{mB}{2}\right)^2-\left(\sqrt{B}\kappa\right)^2\right)\sin^2\theta +(B-1)\kappa^2 \geq 0\ . 
\eeq
Here, we used the conditions (\ref{6-4}) and (\ref{6-6b}). 
It is important to see that the condition (i) holds at any value of $\theta$. 
It may be convenient to treat the condition (ii) by classifying it into two cases (a) and (b): 
\bsub\label{6-8}
\beq
& &{\rm (a)}\ \ \ \left(\frac{mB}{2}\right)^2-\left(\sqrt{B}\kappa\right)^2=0\ , \quad {\rm i.e.,}\quad
|\kappa|=\frac{m}{2}\sqrt{B}\ , 
\label{6-8a}\\
& &{\rm (b)}\ \ \ \left(\frac{mB}{2}\right)^2-\left(\sqrt{B}\kappa\right)^2>0\ , \quad {\rm i.e.,}\quad
|\kappa|<\frac{m}{2}\sqrt{B}\ . 
\label{6-8b}
\eeq
\esub
In the present treatment, $\kappa$ is given as an initial condition and the case (a) gives us time-independent 
$\Lambda^a(=mB/2)$.

%
%%%%%%%%%%%%%%%%%%%%%%%%%%%%%%%%%%%%%%%%%%%%%%%%%%%%%%%%%%%%%%%%%%%%%%
\begin{figure}[t]
\begin{center}
\includegraphics[height=10cm]{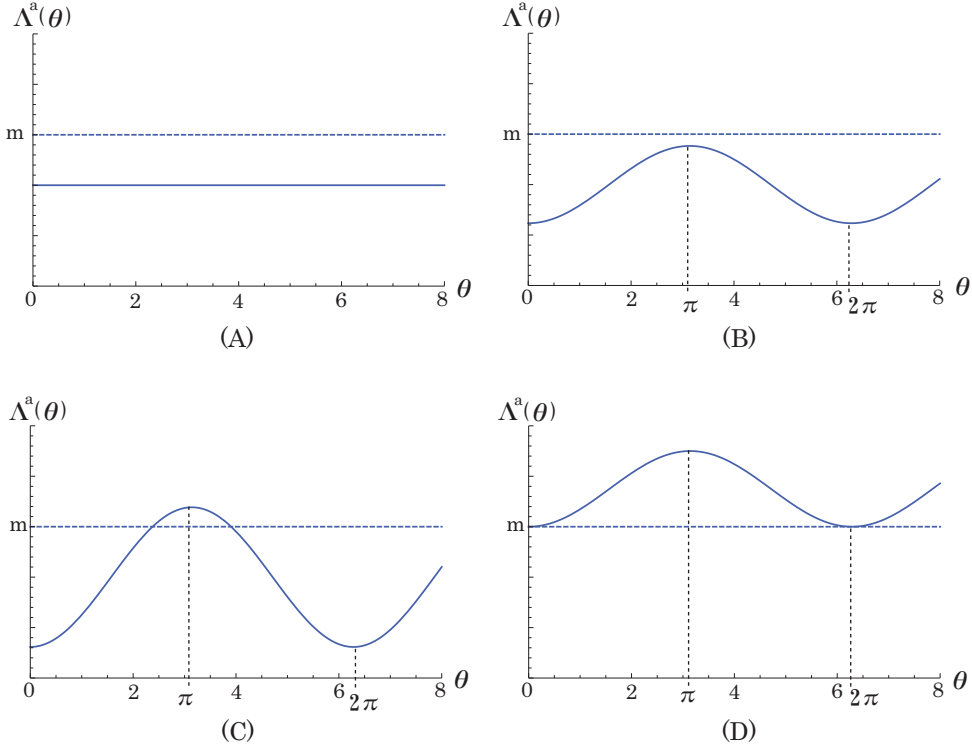}
\caption{Four cases under considerations are schematically depicted.
}
\label{fig:4}
\end{center}
\end{figure}
%%%%%%%%%%%%%%%%%%%%%%%%%%%%%%%%%%%%%%%%%%%%%%%%%%%%%%%%%%%%%%%%%%%%%%%%
%

Consideration on the condition (iii) is rather lengthy and, then, only the results will be presented. 
In this case, it may be successful to consider the problem under the following 
four cases depicted in Fig.\ref{fig:4}: 
The case (A) is nothing but the case (a) given in the relation (\ref{6-8a}). 
Since $\Lambda^a(\tau)\leq m$, we have 
\beq\label{6-9}
1\leq B \leq 2\ , \qquad 
|\kappa|=\frac{m}{2}\sqrt{B}\ , \qquad
\Lambda^a(\tau)=\Lambda^a_A(\tau)=\frac{mB}{2}\ . 
\eeq
In the case (B), the following inequality holds: 
\beq\label{6-10}
\frac{mB}{2}+\sqrt{\left(\frac{mB}{2}\right)^2-\left(\sqrt{B}\kappa\right)^2}\leq m\ . 
\eeq
By solving the inequality (\ref{6-10}), we have 
\beq\label{6-11}
1\leq B < 2\ , \qquad
m\sqrt{1-\frac{1}{B}} \leq |\kappa| <\frac{m}{2}\sqrt{B}\ . 
\eeq
Here, of course, we used the result of the case (b). 
Then, if at the initial time $\tau=0$, $\theta=0$ is chosen, i.e., $\chi_0=0$ in the 
general solution (\ref{5-26}), we have 
\beq\label{6-12}
\Lambda^a(\tau)=\Lambda^a_B(\tau)=\frac{mB}{2}-\sqrt{\left(\frac{mB}{2}\right)^2-\left(\sqrt{B}\kappa\right)^2}\ \cos\left(2\gamma\sqrt{B}\tau\right)\ . 
\eeq

The case (C) satisfies the inequality 
\beq\label{6-13}
\frac{mB}{2}-\sqrt{\left(\frac{mB}{2}\right)^2-\left(\sqrt{B}\kappa\right)^2} < m <
\frac{mB}{2}+\sqrt{\left(\frac{mB}{2}\right)^2-\left(\sqrt{B}\kappa\right)^2} \ . 
\eeq
In this case, we obtain 
\beq\label{6-14}
B>1\ , \qquad |\kappa| < m\sqrt{1-\frac{1}{B}}\ . 
\eeq
We adopt the same initial condition as that in the above, 
$\theta=0$, i.e., $\chi_0=0$ at $\tau=0$. 
As is shown in Fig.\ref{fig:4}, there exists an angle 
$\theta_m$ and it is given in the form 
\beq\label{6-15}
& &\frac{mB}{2}-\sqrt{\left(\frac{mB}{2}\right)^2-\left(\sqrt{B}\kappa\right)^2}\ \cos\theta_m=m\ , \nonumber\\
{\rm i.e.,}\ \ 
& &\cos\theta_m=\frac{m\left(\frac{B}{2}-1\right)}{\sqrt{\left(\frac{mB}{2}\right)^2-\left(\sqrt{B}\kappa\right)^2}}\ , 
\quad
(0 < \theta_m < \pi)\ . 
\eeq 
At the time $\tau_m=\theta_m/(2\gamma\sqrt{B})$, $\Lambda^a(\theta_m)=m$ and in the interval 
$\tau=0 \rightarrow \tau_m$, $\Lambda^a(\tau)$ can be expressed as 
\beq\label{6-16}
\Lambda^a(\tau)=\Lambda^a_0(\tau)=
\frac{mB}{2}-\sqrt{\left(\frac{mB}{2}\right)^2-\left(\sqrt{B}\kappa\right)^2}\ \cos\left(2\gamma\sqrt{B}\tau\right)\ . 
\eeq
However, after $\tau=\tau_m$, $\Lambda^a_0(\tau)$ cannot be adopted, because, 
if it is permitted, $\Lambda^a_0(\tau)>m$. 
Then, we define the following function: 
\beq\label{6-17}
\Lambda^a(\tau)=\Lambda^a_1(\tau)=
\frac{mB}{2}-\sqrt{\left(\frac{mB}{2}\right)^2-\left(\sqrt{B}\kappa\right)^2}\ \cos\left(2\gamma\sqrt{B}(\tau-2\tau_m)\right)\ . 
\eeq
The function $\Lambda^a_1(\tau)$ satisfies $\Lambda_1^a(\tau_m)=\Lambda_0^a(\tau_m)=m$ and in the 
interval $\tau=\tau_m \rightarrow 3\tau_m$, $\Lambda_1^a(\tau)<m$. 
Further, in the interval $\tau=3\tau_m \rightarrow 5\tau_m$, we define 
$\Lambda_2^a(\tau)$ in the form 
\beq\label{6-18}
\Lambda^a(\tau)=\Lambda_2^a(\tau)=
\frac{mB}{2}-\sqrt{\left(\frac{mB}{2}\right)^2-\left(\sqrt{B}\kappa\right)^2}\ \cos\left(2\gamma\sqrt{B}(\tau-4\tau_m)\right)\ . 
\eeq
Certainly, $\Lambda_2^a(3\tau_m)=\Lambda_1^a(3\tau_m)=m$ and $\Lambda_2^a(\tau)$ is useful 
in the interval $\tau=5\tau_m \rightarrow 7\tau_m$. 
By proceeding with this task, we arrive at the following solution: 
\beq\label{6-19}
& &\Lambda^a(\tau)=\Lambda^a_C(\tau)=\Lambda_n^a(\tau)
=
\frac{mB}{2}-\sqrt{\left(\frac{mB}{2}\right)^2-\left(\sqrt{B}\kappa\right)^2}\ \cos\left(2\gamma\sqrt{B}(\tau-2n\tau_m)\right)\ , 
\nonumber\\
& &\quad{\rm for}\quad
(2n-1)\tau_m \leq \tau \leq (2n+1)\tau_m\ .\quad (n=0,\ 1,\ 2,\ 3\, \cdots)
\eeq
In the case (D), we have the relation 
\beq\label{6-20}
\frac{mB}{2}-\sqrt{\left(\frac{mB}{2}\right)^2-\left(\sqrt{B}\kappa\right)^2}=m\ . 
\eeq
%
%%%%%%%%%%%%%%%%%%%%%%%%%%%%%%%%%%%%%%%%%%%%%%%%%%%%%%%%%%%%%%%%%%%%%%
\begin{figure}[t]
\begin{center}
\includegraphics[height=5.0cm]{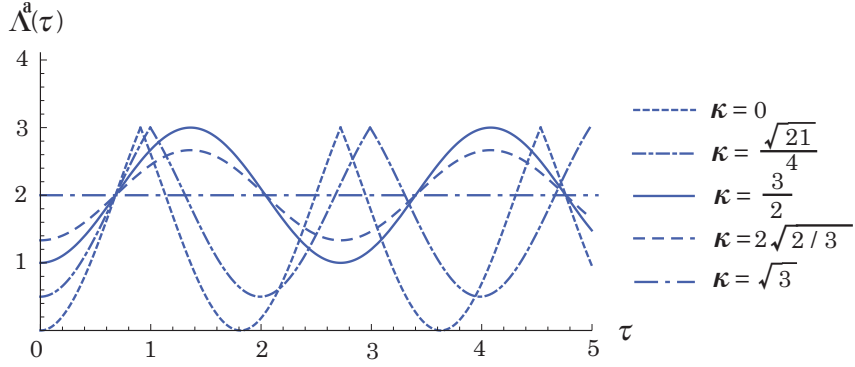}
\caption{The behavior of $\Lambda^a(\tau)$ for various values of $\kappa$ is shown with the same parameters as 
those in Fig.\ref{fig:3} except for $\kappa$. 
}
\label{fig:6}
\end{center}
\end{figure}
%%%%%%%%%%%%%%%%%%%%%%%%%%%%%%%%%%%%%%%%%%%%%%%%%%%%%%%%%%%%%%%%%%%%%%%%
%
%
%
%%%%%%%%%%%%%%%%%%%%%%%%%%%%%%%%%%%%%%%%%%%%%%%%%%%%%%%%%%%%%%%%%%%
\begin{table}[b]
\caption{}
\label{table3}
\centering
\begin{tabular}{c|c|c}%%%The number of columns has to be defined here
\hline
$B$ & $\kappa$ & $\Lambda^a(\tau)$\\
\hline
$B=1$ & $0\leq |\kappa| <\frac{m}{2}$ & $\Lambda^a_B(\tau)$ \\
\cline{2-3} 
 & $|\kappa| =\frac{m}{2}$ & $\Lambda^a_A(\tau)$ \\
\hline
$1<B<2$ & $0\leq |\kappa| <m\sqrt{1-\frac{1}{B}}$ & $\Lambda^a_C(\tau)$ \\
\cline{2-3} 
 & $m\sqrt{1-\frac{1}{B}} \leq |\kappa| <\frac{m}{2}\sqrt{B}$ & $\Lambda^a_B(\tau)$ \\
\cline{2-3} 
 & $|\kappa| =\frac{m}{2}\sqrt{B}$ & $\Lambda^a_A(\tau)$ \\
\hline
$B=2$ & $0\leq |\kappa| <\frac{m}{\sqrt{2}}$ & $\Lambda^a_C(\tau)$ \\
\cline{2-3} 
 & $|\kappa| =\frac{m}{\sqrt{2}}$ & $\Lambda^a_A(\tau)$ \\
\hline
$B>2$ & $0\leq |\kappa| <m\sqrt{1-\frac{1}{B}}$ & $\Lambda^a_C(\tau)$ \\
\cline{2-3} 
 & $|\kappa| =m\sqrt{1-\frac{1}{B}}$ & $\Lambda^a_D(\tau)$ \\
%%%% Table body
\hline
%Parameter set II & 0.800 & 20.0 & 6.6 \\
%\hline
\end{tabular}
\end{table}%%%End of the table
%%%%%%%%%%%%%%%%%%%%%%%%%%%%%%%%%%%%
%
Solution of this equation is given as 
\beq\label{6-21}
B>2\ , \qquad 
|\kappa|=m\sqrt{1-\frac{1}{B}}\ , \qquad
\Lambda^a(\tau)=\Lambda^a_D(\tau)=m\ . 
\eeq
The case (D) is regarded as the limit $\theta_m\rightarrow 0$ in the case (C). 
The function $\Lambda^a_D(\tau)$ does not depend on $\tau$, but its origin is different from that in the case (A). 
Figure \ref{fig:6} shows the behavior of $\Lambda^a(\tau)$ for various values of $\kappa$. 
The same parameters as those used in Fig.\ref{fig:3} are adopted except for $\kappa$ which is a conserved 
quantity determined by the initial condition. 
Under this parameter set, we obtain $B=4/3$ which is in the range $1<B<2$. 
Then, for various values of $\kappa$, the function $\Lambda^a(\tau)$ is turned into $\Lambda_C^a(\tau)$ or  
$\Lambda_B^a(\tau)$ or $\Lambda_A^a(\tau)$ according to Table \ref{table3}. 
For $\kappa=0$ and $\kappa=\sqrt{21}/4\approx 1.1456$, we take $\Lambda^a(\tau)=\Lambda_C^a(\tau)$. 
For $\kappa=3/2$, we adopt $\Lambda^a(\tau)=\Lambda_C^a(\tau)=\Lambda_B^a(\tau)$. 
For $\kappa=2\sqrt{2/3}\approx 1.633$, we chose $\Lambda^a(\tau)=\Lambda_B^a(\tau)$. 
Finally, for $\kappa=\sqrt{3}\approx 1.732$, we adopt $\Lambda^a(\tau)=\Lambda_B^a(\tau)=\Lambda_A^a(\tau)$. 

\begin{figure}[t]
\begin{center}
\includegraphics[height=3.5cm]{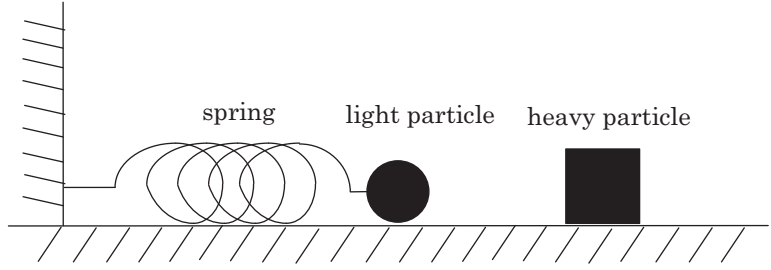}
\caption{The elastic collision of simply oscillating light particle with sufficiently heavy particle is illustrated.
}
\label{fig:5}
\end{center}
\end{figure}

In classical mechanics, we can find the same problem as that discussed in this section: 
elastic collision of simply oscillating light particle with sufficiently heavy particle, 
which is illustrated in Fig.\ref{fig:5}. 
The results obtained in the above are summarized in Table \ref{table3}.
%
%%%%%%%%%%%%%%%%%%%%%%%%%%%%%%%%%%%%%%%%%%%%%%%%%%%%%%%%%%%%%%%%%%%%%%
%%%%%%%%%%%%%%%%%%%%%%%%%%%%%%%%%%%%%%%%%%%%%%%%%%%%%%%%%%%%%%%%%%%%%%%%
%

In this paper, we proposed a new boson representation of the $su(2)$-algebra. 
The basic idea comes from the pseudo $su(1,1)$-algebra in the Schwinger boson 
representation. 
In a certain sense, ours is on the opposite side of the Schwinger representation 
of the $su(2)$-algebra. 
In next paper, Part II, we will prove that ours satisfies the $su(2)$-algebra 
in the subspace (\ref{2-8}) of the whole space (\ref{2-5}) for the case $t_m=C_m+1-t$.

\section*{Acknowledgment}

One of the authors (M.Y.) would like to express his sincere thanks to Mrs. Y. Miyamoto 
for her cordial encouragement.  
One of the authors (Y.T.) is partially supported by the Grants-in-Aid of the Scientific Research 
(No.23540311, No.26400277) from the Ministry of Education, Culture, Sports, Science and 
Technology in Japan.


%\bibliographystyle{ptephy}
%\bibliography{sample}
%
% once the .bbl file has been generated then place the text in your article.

%\vfill\pagebreak

\end{document}